\documentclass[aps,pre,reprint,superscriptaddress,longbibliography,nofootinbib]{revtex4-2}

\usepackage{graphicx}
\usepackage{amsmath,amssymb,bm}
\usepackage{booktabs}
\usepackage{dcolumn}
\usepackage{hyperref}
\usepackage{xcolor}

\hypersetup{
  colorlinks=true,
  linkcolor=blue,
  citecolor=blue,
  urlcolor=blue
}

\newcommand{\figpanel}[2]{\includegraphics[width=#1\textwidth]{#2}}

\begin{document}

\title{Conditioning on a Volatility Proxy Compresses the Apparent Timescale of Collective Market Correlation}

\author{Yuda Bi}
\email{ybi3@gsu.edu} 
\affiliation{Center for Translational Research in Neuroimaging and Data Science (TReNDS), Atlanta, Georgia 30303, USA}

\author{Vince D.\ Calhoun}
\affiliation{Center for Translational Research in Neuroimaging and Data Science (TReNDS), Atlanta, Georgia 30303, USA}
\affiliation{Georgia State University, Georgia Institute of Technology, and Emory University, Atlanta, Georgia 30303, USA}

\date{\today}

\begin{abstract}
We address the attribution problem for apparent slow collective dynamics:
is the observed persistence intrinsic, or inherited from a persistent
driver? For the leading eigenvalue fraction
$\psi_1=\lambda_{\max}/N$ of S\&P 500 60-day rolling correlation matrices
($237$ stocks, 2004--2023), a VIX-coupled Ornstein--Uhlenbeck model
reduces the effective relaxation time from $298$ to $61$ trading days and
improves the fit over bare mean reversion by $\Delta$BIC$=109$. On the
decomposition sample, an informational residual of $\log(\mathrm{VIX})$
alone retains most of that gain ($\Delta$BIC$=78.6$), whereas a mechanical
VIX proxy alone does not improve the fit. Autocorrelation-matched placebo
fields fail ($\Delta$BIC$_{\max}=2.7$), disjoint weekly reconstructions still
favor the field-coupled model ($\Delta$BIC$=140$--$151$), and six anchored
chronological holdouts preserve the out-of-sample advantage. Quiet-regime and
field-stripped residual autocorrelation controls show the same collapse of
persistence. Stronger hidden-variable extensions remain only partially
supported. Within the tested stochastic class, conditioning on the observed
VIX proxy absorbs most of the apparent slow dynamics.
\end{abstract}

\maketitle

\section{Introduction}

\subsection{Slow collective observables pose an attribution problem}

Slow collective observables are routinely interpreted as evidence for
intrinsic memory, metastability, or critical slowing down in complex systems
\cite{Scheffer2009,Allen2014,Friedrich2011}. The same empirical appearance
can, however, arise when a moderately relaxing observable tracks an
equilibrium that is itself displaced by a persistent external or latent field
\cite{Zwanzig2001,Risken1996,Gardiner2009,Falkena2019}. Distinguishing these
two mechanisms is difficult whenever the field cannot be turned on and off
experimentally, which is precisely the situation in ecological synchrony,
dynamic brain connectivity, and climate variability
\cite{Scheffer2009,Allen2014,Falkena2019}. The first logical task is
therefore attribution: before positing intrinsic memory, one should ask
whether the apparent persistence is inherited from the motion of a driver
\cite{Zwanzig2001,Theiler1992,Gardiner2009}.

That attribution problem is sharper than a generic model-selection exercise
\cite{White1982,KassRaftery1995}. A slow observable can often be fit
acceptably by several low-dimensional descriptions, including bare mean
reversion, multistability, hidden-variable models, and explicit field-coupled
dynamics \cite{Risken1996,Gardiner2009,Scheffer2009}. Good in-sample fit
alone therefore does not establish mechanism \cite{White1982}. What is needed
is a sequence of discriminations that asks which parts of the observed
persistence survive conditioning, placebo substitution, and model-free
controls \cite{Theiler1992,Zwanzig2001,Falkena2019}. The present manuscript
is built around that sequence.

\subsection{Financial correlation dynamics provide a stringent test case}

Financial markets offer an unusually demanding testbed because they provide
long, high-dimensional, and economically interpretable correlation time series
\cite{Laloux1999,Plerou1999,Plerou2002,Bun2017,Potters2020,Livan2018}. Daily
correlation matrices of large-cap U.S. equities exhibit a dominant market mode
that is stable enough to track over decades and strong enough to matter for
systemic-risk interpretation
\cite{Laloux1999,Plerou2002,Billio2012,Kritzman2011}. At the same time, the
candidate field proxy is not cleanly exogenous: volatility indices, realized
co-movement, and broad risk sentiment are partly entangled by construction
\cite{ForbesRigobon2002,Bekaert2005,Cont2001,Engle2002}. If an attribution
workflow works here, it is likely to transfer to cleaner domains rather than
only to easier ones.

The target of inference is therefore conditional attribution with respect to
an observed field proxy, not proof of direct exogenous forcing
\cite{White1982,Zwanzig2001}. That distinction matters in finance because the
strongest plausible confound is a latent common driver that moves both VIX and
the collective correlation observable \cite{ForbesRigobon2002,Cont2001}. The
present design is built to separate intrinsic persistence, persistence-only
surrogates, and purely mechanical overlap from that broader class of
field-proxy explanations, while remaining explicit about what direct-coupling
claims it cannot settle \cite{White1982,Theiler1992}.

The financial literature also supplies sharply competing narratives
\cite{Sornette2003,LuxMarchesi1999,ContBouchaud2000,Bouchaud2013,MarsiliRaffaelli2006}.
One strand emphasizes endogenous collective reorganization, cascade
amplification, and near-critical dynamics
\cite{Sornette2003,LuxMarchesi1999,ContBouchaud2000,Bak1987,Sethna2001}.
Another strand emphasizes crisis-period heteroskedasticity, identifiable
macro-financial shocks, and volatility-driven co-movement
\cite{Cont2001,ForbesRigobon2002,Bekaert2005,Engle2002,Billio2012}. These
views need not be mutually exclusive, but they imply different expectations
for the shape of the effective dynamics: intrinsic criticality points toward
persistent double-well or memory-bearing structure, whereas field conditioning
points toward a comparatively simple observable that relaxes around a moving
equilibrium \cite{Risken1996,Gardiner2009,Scheffer2009}. That distinction is
exactly what the present tests are designed to probe.

Within that setting we study the leading-eigenvalue fraction
\begin{equation}
\psi_1(t)=\frac{\lambda_{\max}(t)}{\mathrm{tr}\,C(t)}
=\frac{\lambda_{\max}(t)}{N},
\end{equation}
where $C(t)$ is the rolling Pearson correlation matrix of stock returns. This
quantity inherits the main intuition of random-matrix studies of financial
correlation, namely that the dominant eigenmode captures collective market-wide
alignment while the bulk largely reflects noise and sectoral substructure
\cite{Laloux1999,Plerou1999,Plerou2002,Marchenko1967,Bun2017}. It is also
closely related to the absorption ratio and to state-classification approaches
based on the correlation spectrum
\cite{Kritzman2011,Munnix2012,Pharasi2018,Pharasi2020}. What changes here is
not the spectral object, but the question: rather than using the leading mode
as a static diagnostic, we treat it as a dynamical variable whose persistence
itself must be explained \cite{Tumminello2010,Fenn2011,Kenett2010,Song2011}.
This choice is useful precisely because $\psi_1$ remains collective enough to
track systemic episodes while remaining simple enough to admit explicit
stochastic competition between field-free, field-coupled, and hidden-variable
models \cite{Kritzman2011,Munnix2012,Billio2012,Risken1996,Gardiner2009,White1982}.

\subsection{The literature gap is not whether memory exists, but what causes it}

The immediate empirical background is the generalized-Langevin work of Wand,
He{\ss}ler, and Kamps on the mean market correlation of the S\&P 500
\cite{Wand2023Entropy,Hessler2023}. Their results are important because they
show that a one-dimensional memory-bearing description of collective market
correlation is empirically viable and that the inferred memory kernel is not
negligible \cite{Wand2023Entropy,Wand2023Arxiv}. More broadly, they connect
financial correlation dynamics to a growing literature on projected memory
kernels and hidden slow variables in complex systems
\cite{Zwanzig2001,Falkena2019,Rinn2015,Stepanov2015}. But that approach leaves
open a central causal question: does the apparent memory remain once explicit
external fields are introduced and compared against persistence-matched placebo
drivers \cite{Theiler1992,Zwanzig2001}?

That gap matters because several adjacent literatures already suggest that
volatility conditions strongly modulate correlations during crises
\cite{ForbesRigobon2002,Bekaert2005,Cont2001,Engle2002}. The options-implied
volatility index VIX is especially relevant because it is persistent,
forward-looking, and economically tied to broad uncertainty shocks
\cite{Bloom2009,Cont2001,Bekaert2005}. At the same time, VIX is partly
mechanical because both VIX and realized correlation respond to the same
underlying return covariance matrix \cite{ForbesRigobon2002,Cont2001}. A
serious argument for field-driven dynamics must therefore do more than report
a high correlation between VIX and $\psi_1$. It must separate informational
from mechanical overlap, show that persistence-matched surrogates fail, and
test whether stronger hidden-variable interpretations still add anything after
the one-dimensional field problem has been settled
\cite{Theiler1992,White1982,Wand2023Entropy}.

Even then, a positive result against VIX should be read as evidence that an
observed field proxy carries structured coupling information, not as proof
that the proxy is the uniquely causal driver
\cite{White1982,Zwanzig2001,ForbesRigobon2002}. That is the inferential level
at which the paper's main claim is framed.

\subsection{This paper provides a layered attribution test}

The manuscript proceeds in three deliberately unequal layers. The first layer
is the main result: we compare explicit one-dimensional stochastic models and
show that conditioning on an observed VIX field proxy absorbs most of the
apparent persistence in $\psi_1$ within the tested stochastic class, while
autocorrelation-matched placebo fields fail by a wide margin. The second layer
asks a harder question suggested by the
generalized-Langevin literature: can the observed VIX itself serve as the
hidden coordinate of an exact two-dimensional linear-Gaussian system whose
projection generates memory
\cite{Zwanzig2001,Wand2023Entropy,Wand2023Arxiv,Falkena2019}? The third layer
tests whether an orthogonal residual beyond VIX functions as a meaningful
secondary order parameter or merely as a descriptive leftover
\cite{Driessen2009,White1982}.

The evidential profile is intentionally asymmetric. The one-dimensional
field-proxy attribution is strongly supported by formal model comparison,
placebo rejection, directionality diagnostics, field decomposition, and
model-free persistence collapse. The two-dimensional bridge survives only
partially: it becomes more plausible under a Wand-faithful weekly
construction, but it does not recover the reported memory scale cleanly. The
orthogonal residual is descriptively nontrivial, but its forecasting role is
not supported. This asymmetry is a feature rather than a flaw, because it
keeps the strongest claim tied to the strongest evidence
\cite{KassRaftery1995,White1982}.

The broader methodological point is transferable beyond finance. Any system
that supplies a collective observable and a candidate field can be subjected
to the same logic: build the collective variable, compare bare and
field-coupled stochastic models, reject persistence-only surrogates in the
spirit of surrogate-data testing, and only then move to stronger
hidden-variable or residual-state interpretations
\cite{Theiler1992,Allen2014,Falkena2019,Gilson2023}. The rest of the paper
follows that order. Section II details the data, experimental design, and
estimation procedures used throughout. Section III presents the
one-dimensional attribution result and its robustness tests, then the
two-dimensional extension, and finally the orthogonal residual analysis.
Section IV discusses what these results do and do not imply for collective
market dynamics and for the broader study of apparent slow modes
\cite{Scheffer2009,Zwanzig2001,Wand2023Entropy}.

\section{Methods}

\subsection{Data, alignment, and construction of the collective observable}

The daily baseline uses 237 S\&P 500 constituents with continuous
adjusted-close coverage from 2004 through 2023. Daily log returns are
computed in the usual way,
$r_i(t)=\log P_i(t)-\log P_i(t-1)$, and 60-trading-day rolling Pearson
correlation matrices are formed from those returns, following the standard
correlation-matrix construction used throughout the financial random-matrix
literature \cite{Laloux1999,Plerou1999,Plerou2002,Bun2017}. The collective
observable is the leading-eigenvalue fraction
\begin{equation}
\psi_1(t)=\frac{\lambda_{\max}(t)}{\mathrm{tr}\,C(t)}
=\frac{\lambda_{\max}(t)}{N},
\end{equation}
where $\mathrm{tr}\,C=N$ because $C$ is a Pearson correlation matrix with unit
diagonal \cite{Laloux1999,Plerou2002,Potters2020}. This observable is the
dynamic analogue of the dominant market mode emphasized in random-matrix and
absorption-ratio studies, but here it is treated as the dependent variable of
a stochastic model rather than as a descriptive summary statistic
\cite{Kritzman2011,Munnix2012,Pharasi2018,Pharasi2020}.

Daily VIX closing values are aligned to the same trading calendar using
Federal Reserve Economic Data. After alignment, the one-dimensional SDE sample
contains $n=4973$ daily observations from 2004-03-30 through 2023-12-29. On
that aligned sample, $\psi_1$ has mean $0.3778$, standard deviation $0.1171$,
and maximum $0.7518$ during the COVID-19 episode. These levels are not
interpreted as thermodynamic invariants; they depend in part on the finite
rolling-window estimator and on the high-dimensional aspect ratio of the
correlation matrices \cite{Marchenko1967,Bun2017,Potters2020}.

The ratio $W/N=60/237\approx 0.25$ places each rolling correlation matrix in a
regime where the Marchenko--Pastur bulk edge is nontrivial
\cite{Marchenko1967,Bun2017}. That fact matters for level interpretation
because the baseline value of $\psi_1$ contains a stationary noise-floor
contribution from finite-sample eigenvalue spreading \cite{Bun2017,Potters2020}.
It matters less for the present attribution exercise because the paper focuses
on temporal comparisons---autocorrelation structure, model discrimination, and
field conditioning---rather than on reading the absolute level of $\psi_1$ as
a free-energy-like order parameter \cite{Livan2018,Potters2020,White1982}. The
same construction is recomputed directly from raw returns for
$W=30,45,60,90,120$ in the window-robustness experiment reported below and in
Appendix~\ref{sec:app_tables}.

\subsection{Experimental design and evidential logic}

The empirical design is sequential rather than omnibus. Each stage is tied to
a distinct rival explanation for the observed slowness of $\psi_1$
\cite{White1982,Theiler1992,Zwanzig2001}. The primary comparison is between a
bare OU model, which attributes persistence to the observable itself, and a
VIX-coupled OU model, which attributes part of that persistence to motion of a
field-dependent equilibrium \cite{Risken1996,Gardiner2009,Cont2001}. Quartic
and regime-switching alternatives are then added to test whether the data
favor endogenous multistability or rare discrete stress states over the
continuous-field picture \cite{Sornette2003,Hamilton1989,AngBekaert2002}.

The main causal threat is that any sufficiently persistent driver might look
good once inserted into the likelihood. We therefore define a formal placebo
gate in which the real VIX field must beat 100 persistence-matched surrogate
fields generated from the fitted marginal dynamics of $\log(\mathrm{VIX})$
\cite{Theiler1992}. A second threat is circularity: VIX and $\psi_1$ are both
functions of the same return covariance matrix. To address that concern we
decompose VIX into mechanical and informational components, fit each component
as a standalone field, and compare their model-selection power directly
\cite{ForbesRigobon2002,Cont2001,White1982}.

The design then moves outside the likelihood. Quiet-regime and field-stripped
autocorrelation tests ask whether persistence still collapses when the field is
approximately dormant or when the fitted conditional equilibrium is removed.
The exact window-size sweep asks whether the field effect survives changes in
time resolution. Disjoint weekly reconstructions and same-horizon disjoint
60-day blocks ask whether the core one-dimensional attribution survives once
measurement overlap is reduced or removed. Anchored chronological holdouts ask
whether the main result is tied to one historical partition. MOVE and TED
controls ask whether the field picture is broader than VIX, while the exact
two-dimensional model asks whether the observed VIX can itself close the
hidden-variable bridge
suggested by generalized-Langevin work
\cite{Wand2023Entropy,Wand2023Arxiv,Falkena2019}. Finally, the orthogonal
residual analysis tests whether a VIX-orthogonal state variable carries
operational information beyond the main field story \cite{Driessen2009}.

This layered design is deliberate. The paper's strongest claim is attached to
the largest evidential stack, not to the most ambitious theoretical extension
\cite{KassRaftery1995,White1982}. The robustness procedures are specified here
and written out in execution detail in Appendix~\ref{sec:app_robustness} so
that the paper reads as a tightly linked sequence of discrimination tests
rather than as a collection of disconnected checks.

The design also has a clear inference boundary. It can strongly distinguish
bare intrinsic persistence from conditioning on an observed field proxy, and
it can separately test autocorrelation-only and mechanical-overlap objections.
It cannot, by itself, distinguish direct coupling from a latent common driver
that moves both the proxy and the observable
\cite{White1982,Zwanzig2001,ForbesRigobon2002}.

\begin{table*}
\caption{\label{tab:logic}Layered attribution logic. The strongest claim is the
one that survives the largest stack of discrimination tests.}
\centering
\small
\begin{tabular}{lll}
\toprule
\parbox[t]{0.24\textwidth}{\raggedright Rival explanation} &
\parbox[t]{0.49\textwidth}{\raggedright Primary discrimination} &
\parbox[t]{0.18\textwidth}{\raggedright Outcome in this manuscript} \\
\midrule
\parbox[t]{0.24\textwidth}{\raggedright Bare intrinsic persistence or autocorrelation-only inheritance} &
\parbox[t]{0.49\textwidth}{\raggedright M0 versus M2, placebo surrogates, quiet-regime ACF collapse, field-stripped residual ACF, disjoint weekly reconstructions, and anchored holdouts} &
\parbox[t]{0.18\textwidth}{\raggedright Strongly disfavored} \\
\parbox[t]{0.24\textwidth}{\raggedright Purely mechanical overlap between VIX and $\psi_1$} &
\parbox[t]{0.49\textwidth}{\raggedright Mechanical/informational decomposition, standalone field tests, and recipe sensitivity across freeze and weighting choices} &
\parbox[t]{0.18\textwidth}{\raggedright Strongly disfavored dynamically} \\
\parbox[t]{0.24\textwidth}{\raggedright Observed VIX as the exact hidden coordinate behind projected memory} &
\parbox[t]{0.49\textwidth}{\raggedright Exact two-dimensional linear-Gaussian comparison and projected-kernel timescale check} &
\parbox[t]{0.18\textwidth}{\raggedright Only partially supported} \\
\parbox[t]{0.24\textwidth}{\raggedright VIX-orthogonal residual as an operational predictor} &
\parbox[t]{0.49\textwidth}{\raggedright Q2-versus-Q3 future-VIX test across 30-, 60-, and 90-day horizons} &
\parbox[t]{0.18\textwidth}{\raggedright Not supported} \\
\bottomrule
\end{tabular}
\end{table*}

\subsection{One-dimensional stochastic hierarchy and exact daily likelihood}

The core model comparison is intentionally narrow. We test whether the daily
dynamics of $\psi_1$ are better described by bare mean reversion, by mean
reversion around a field-dependent equilibrium, or by simple nonlinear
alternatives often associated with metastability or effective double wells
\cite{Risken1996,Gardiner2009,Scheffer2009}. The two central models are
\begin{align}
d\psi_1 &= -\theta_0(\psi_1-\mu_0)\,dt+\sigma_0\,dW_t,\\
d\psi_1 &= \left[-\theta(\psi_1-\mu)+\beta v_t\right]dt+\sigma\,dW_t,
\end{align}
with $v_t=\log(\mathrm{VIX}_t)$. The first model, M0, attributes all
persistence to the observable itself. The second model, M2, allows the
equilibrium to slide with the volatility field, so that
$\mu_{\mathrm{eff}}(v)=\mu+(\beta/\theta)v$
\cite{Risken1996,Gardiner2009,Cont2001}. We also fit quartic-drift variants
with and without the field, plus constrained two-state regime-switching
competitors used only to calibrate whether a rare extreme-state channel adds
structure beyond the continuous field
\cite{Hamilton1989,AngBekaert2002,Sornette2003}.

For M0 and M2, the exact one-step Gaussian transition law is available in
closed form because the drift is linear conditional on the field path.
Treating $v_t$ as piecewise constant over each daily interval, the exact daily
transition is
\begin{equation}
\psi_{1,t+1}\mid \psi_{1,t},v_t \sim \mathcal{N}\!\left(m_t,\;q\right),
\end{equation}
with
\begin{align}
m_t &= e^{-\theta}\psi_{1,t}
 + \left(1-e^{-\theta}\right)
\left(\mu+\frac{\beta}{\theta}v_t\right),\\
q &= \frac{\sigma^2}{2\theta}\left(1-e^{-2\theta}\right),
\end{align}
and the obvious M0 specialization $\beta=0$
\cite{Risken1996,Gardiner2009}. The exact derivation is given in
Appendix~\ref{sec:app_exact}. We maximize the resulting Gaussian likelihood
directly for M0 and M2. Quartic variants are optimized numerically under the
same daily increment convention, and the constrained regime-switching models
are estimated with a Gaussian Hamilton filter under a two-state Markov
transition matrix \cite{Hamilton1989,AngBekaert2002}.

Model comparison is based on BIC, with AIC recorded as a secondary diagnostic.
Because all fitted models are coarse-grained approximations to a far richer
market process, the preferred specification is interpreted in the pseudo-true
sense of misspecified likelihood theory rather than as a literal
data-generating law \cite{White1982}. This point matters for the timescale
language used later: the reported $\tau_{\mathrm{auto}}=1/\theta_0$ and
$\tau_{\mathrm{cond}}=1/\theta$ are model-implied effective relaxation rates
within the fitted OU class, not claims about a unique microscopic law
\cite{White1982,Gardiner2009}. We also test a heteroskedastic variant M2$'$,
defined by the same OU+field drift as M2 but with
\begin{equation}
\sigma(\psi_1)=\sigma_0+\sigma_1\psi_1.
\end{equation}
Because the diffusion coefficient is then state dependent, M2$'$ is fit by
direct numerical likelihood optimization under the same daily increment
convention. Its role is diagnostic rather than headline: it asks whether
modest state-dependent noise changes the field-attribution picture
\cite{Gardiner2009,White1982}. Throughout the paper we summarize the
model-based field contribution by
\begin{equation}
\mathrm{SCPA}=1-\frac{\tau_{\mathrm{cond}}}{\tau_{\mathrm{auto}}}
=1-\frac{\theta_0}{\theta},
\end{equation}
which compares the conditioned and bare OU timescales directly. We use the
usual Jeffreys-style BIC scale only as an evidence summary, not as a
substitute for robustness analysis \cite{KassRaftery1995}. The full execution
details for the model-comparison pipeline are given in
Appendix~\ref{sec:app_robustness}.

\subsection{Placebo gate and the formal attribution null}

The strongest falsification device is a placebo-field gate modeled on
surrogate-data logic \cite{Theiler1992}. We fit an AR($p$) model to
$v_t=\log(\mathrm{VIX}_t)$, select $p$ by AIC over $p=1,\ldots,10$, and
obtain $p=9$ on the aligned sample. We then simulate 100 independent
surrogate fields from that fitted marginal process and rescale each surrogate
to the empirical mean and variance of the real series. Each surrogate then
replaces the real VIX field in the same M2 likelihood. The formal null
hypothesis is that the real VIX improvement can be matched by an independent
realization drawn from the fitted AR marginal law of $\log(\mathrm{VIX})$.
Under that null, any large BIC gain would be attributable to generic
persistence alone rather than to specific coupling information
\cite{Theiler1992,White1982}. Rejection therefore means more than ``VIX is
persistent'': it means that the observed driver carries coupling structure
beyond its own autocorrelation profile.

This placebo design is deliberately conservative. It does not test against
white noise or against an unrelated macro covariate, both of which would be
easy nulls. Instead it tests against fields with essentially the same
low-order temporal persistence as the real series, which is the relevant
adversarial explanation for any claim about inherited persistence
\cite{Theiler1992,Zwanzig2001}. The empirical $p$-value is the fraction of
placebo gains that equal or exceed the real $\Delta$BIC improvement. Because
only 100 placebos are used, the resolution is coarse, but the observed gap is
so large that this discreteness is immaterial for the paper's conclusions.
The full placebo protocol is written out in
Appendix~\ref{sec:app_placebo_protocol}.

\subsection{Directional diagnostics and mechanical versus informational overlap}

Directionality is evaluated with manual bivariate Granger systems in both
levels and first differences. The level test checks whether lagged VIX terms
improve prediction of $\psi_1$ and vice versa. The differenced test is
reported because near-integration can distort standard causality inference
when persistent series are used in levels \cite{TodaYamamoto1995}. These
Granger results are not treated as the primary identification device because
rolling-window overlap and mild misspecification can still affect them, but
they provide a useful directional sanity check alongside the placebo gate
\cite{TodaYamamoto1995,White1982}.

Mechanical overlap is estimated by constructing a VIX proxy that preserves the
empirical correlation path while suppressing time variation in per-stock
volatilities. Concretely, for each stock $i$ we compute its 60-day rolling
volatility series, take the median of that series over the aligned sample,
replace the time-varying volatility by that stock-specific median, and
recompute an equal-weight portfolio variance using the actual rolling
correlation matrices. The square root of that portfolio variance, after a
single multiplicative rescaling to match the unconditional VIX level, defines
the mechanical volatility proxy. This construction is not intended to
reproduce the full option-integral definition of VIX; it is a controlled
decomposition designed to ask how much of the $\psi_1$ association survives
after the common realized-covariance channel is stripped down to its
correlation component \cite{ForbesRigobon2002,Bekaert2005,Cont2001}.

The informational component is then defined as the regression residual of
actual $\log(\mathrm{VIX})$ on the mechanical proxy. The mechanical and
informational fractions reported in the Results section come from a sequential
$R^2$ decomposition: first regress $\psi_1$ on the mechanical proxy alone,
then add the informational residual and record the incremental gain. The same
two components are also inserted separately into the M2 likelihood, producing
standalone field tests that directly assess whether model-selection power
lives in the mechanical channel, the informational channel, or both. The
static variance split is operational rather than unique, so we also repeat the
construction under alternative volatility-freezing choices and portfolio
weights. The point of those variants is not to force one canonical
mechanical/informational percentage, but to test whether the sign and
magnitude of the standalone field gains are recipe dependent
\cite{White1982,ForbesRigobon2002}. The exact algebra used for this
decomposition is written out in
Appendix~\ref{sec:app_exact}, and the full execution protocol is given in
Appendix~\ref{sec:app_mech_protocol}.

\subsection{Model-free persistence controls, non-overlapping reconstructions, and out-of-sample evaluation}

The main one-dimensional claim is intentionally checked with controls that do
not depend on the OU likelihood. The first is a quiet-regime autocorrelation
comparison. Our strict ex ante criterion is daily VIX confined to $[15,18]$
for at least 120 consecutive trading days. Because that criterion yields no
qualifying segment in 2004--2023, we adopt a transparent fallback based on a
20-day rolling-median VIX in $[14,20]$ and then repeat the calculation on
neighboring bands $[13,21]$ and $[15,19]$. For each accepted set of quiet
segments, the autocorrelation function of $\psi_1$ is pooled across segments
and summarized by the e-folding lag, defined as the first lag at which the
ACF falls below $e^{-1}$ of its lag-zero value. This diagnostic asks whether
persistence collapses when the field is approximately quiescent, without
assuming any specific stochastic model.

The second model-free control is the field-stripped residual
\begin{equation}
\epsilon_{\mathrm{M2}}(t)=\psi_1(t)-\mu_{\mathrm{eff}}(v_t),
\end{equation}
where $\mu_{\mathrm{eff}}(v_t)=\mu+(\beta/\theta)v_t$ is the fitted
conditional equilibrium of M2. The ACF of $\epsilon_{\mathrm{M2}}$ is
compared with the ACF of raw $\psi_1$ through both e-folding lags and
integrated ACF mass over fixed horizons. This control differs conceptually
from the OU timescale ratio. The timescale ratio asks how fast the model
relaxes conditional on the field; the residual ACF asks how much temporal
structure remains after subtracting the fitted equilibrium path. The
distinction is important enough that the comparison is revisited explicitly in
the Discussion section. Full execution details for these controls are given in
Appendix~\ref{sec:app_acf_protocol}.

A deeper measurement concern is overlapping windows. We therefore rebuild the
one-dimensional comparison on disjoint weekly observables using 5-trading-day
correlation windows sampled every 5 trading days, once for a weekly $\psi_1$
proxy and once for weekly mean market correlation. To probe the same-horizon
extreme, we also construct disjoint 60-day block versions of $\psi_1$ and pair
each block with either its end-of-block VIX value or its within-block mean VIX
\cite{Wand2023Entropy,Wand2023Arxiv}. These series are lower power than the
daily baseline, but they directly test whether the field-coupled model remains
preferred once overlap is reduced or removed at the observable-construction
stage. Their full execution details are also given in
Appendix~\ref{sec:app_window_protocol}.

Out-of-sample stability is assessed at two levels. The baseline split remains
2016-01-01: parameters are estimated on the pre-2016 observations and
evaluated on the post-2016 observations under the same exact one-step Gaussian
likelihood. Beyond that baseline, we run an anchored split sweep at
2010-01-01, 2012-01-01, 2014-01-01, 2016-01-01, 2018-01-01, and 2020-01-01,
each time refitting on the available prefix and evaluating on the full
remaining suffix. Performance is summarized by average log likelihood per
observation, the test/train ratio, and the test-set likelihood gap between M2
and M0. This remains a chronological rather than cross-validated design, but
it prevents the stability claim from hinging on a single historical partition
\cite{White1982}. The chronological holdout and anchored-sweep protocols are
given in Appendix~\ref{sec:app_window_protocol}.

\subsection{Window-size robustness and resolution dependence}

The window-robustness experiment is designed to answer a specific attack:
whether the baseline conditioned timescale near 61 trading days is merely a
restatement of the 60-day correlation window. We therefore recompute $\psi_1$
directly from raw returns for $W=30,45,60,90,120$, refit M0 and M2 at each
window, and record four quantities for every $W$: the bare timescale
$\tau_0=1/\theta_0$, the conditioned timescale $\tau_{\mathrm{cond}}=1/\theta$,
the susceptibility $\beta/\theta$, and the scalar persistence-collapse
attribution
\begin{equation}
\mathrm{SCPA}(W)=1-\frac{\tau_{\mathrm{cond}}(W)}{\tau_0(W)}.
\end{equation}
The point of this sweep is not to demand that absolute timescales be
identical across resolutions. Rolling correlation estimation is itself a
low-pass filter, so absolute timescales can change mechanically with $W$. The
relevant structural question is instead whether the field effect---as
summarized by the susceptibility and by the sign and magnitude of SCPA---survives
across resolutions. The quantitative results of that sweep are reported in
Appendix Table~\ref{tab:window_robustness_appendix}, and the full execution
protocol is given in Appendix~\ref{sec:app_window_protocol}.

\subsection{Exact two-dimensional extension and Wand-faithful weekly reconstruction}

To test whether the observed VIX can itself play the role of the hidden
variable behind apparent one-dimensional memory, we fit the exact
linear-Gaussian system
\begin{align}
d\psi_1 &= \left[-\theta_{\psi}(\psi_1-\mu_{\psi})+\beta_{\psi}v\right]dt
 + \sigma_{\psi}dW_1,\\
dv &= \left[-\theta_v(v-\mu_v)+\beta_v\psi_1\right]dt + \sigma_v dW_2,
\end{align}
with $v=\log(\mathrm{VIX})$ and correlated innovations. In discrete time this
is an exact VAR(1) transition model whose restriction pattern determines
whether the drift is decoupled, feedforward, or bidirectional
\cite{Gardiner2009,Zwanzig2001}. The fitted discrete transition matrix is
mapped back to continuous time by matrix logarithm, and the innovation
covariance is matched through the discrete Lyapunov equation, yielding exact
continuous-time parameters within the linear-Gaussian class. The
projected-memory derivation associated with this system is given in
Appendix~\ref{sec:app_exact}.

The daily comparison alone is not enough because overlapping 60-day rolling
windows can accentuate serial dependence. To compare more fairly with Wand
\textit{et al.}, we therefore reconstruct disjoint weekly observables using
5-trading-day correlation windows sampled every 5 trading days
\cite{Wand2023Entropy,Wand2023Arxiv}. This reconstruction is performed twice:
once for a weekly $\psi_1$ proxy computed from each disjoint weekly matrix,
and once for the mean market correlation of the same matrices. Both weekly
series are passed through the same exact two-dimensional model comparison and
projected-kernel calculations. This design keeps the hidden-variable test
close to the preprocessing assumptions of the generalized-Langevin literature
while preserving direct comparability with the daily model family.
Full execution details for the 2D comparison are given in
Appendix~\ref{sec:app_twod_protocol}.

\subsection{Orthogonal residual state variable}

The final construction is a level-orthogonal residual beyond the fitted
conditional equilibrium. We estimate
\begin{equation}
\epsilon_{\perp}(t)=\psi_1(t)-\left[a+b\,\log(\mathrm{VIX}(t))\right],
\end{equation}
using ordinary least squares on levels, so that $\epsilon_{\perp}$ is
approximately uncorrelated with $\log(\mathrm{VIX})$ by construction. This
object is not the same as $\epsilon_{\mathrm{M2}}$. The latter strips the
fitted M2 equilibrium and is useful for persistence attribution; the former
strips only the best linear level relation and is useful for asking whether
an additional VIX-orthogonal state variable carries descriptive or predictive
structure. We then partition the $(\log\mathrm{VIX},\epsilon_{\perp})$ plane
by the sample median of $\log\mathrm{VIX}$ and by the sign of
$\epsilon_{\perp}$, producing the four quadrants used in the final results
section.

The proposed predictive test compares future VIX changes after Q2 days
(low VIX, positive $\epsilon_{\perp}$) against Q3 days
(low VIX, negative $\epsilon_{\perp}$). We report mean future changes,
Mann--Whitney $p$-values, and rank-biserial effect sizes at 30-, 60-, and
90-trading-day horizons. This is intentionally a narrow operational claim. A
residual can be dynamically nontrivial without qualifying as a forecasting
signal, and the analysis is designed to keep those two propositions separate
\cite{White1982,Driessen2009}.

\section{Results}

\subsection{Empirical target}

\begin{figure*}
\centering
\figpanel{0.98}{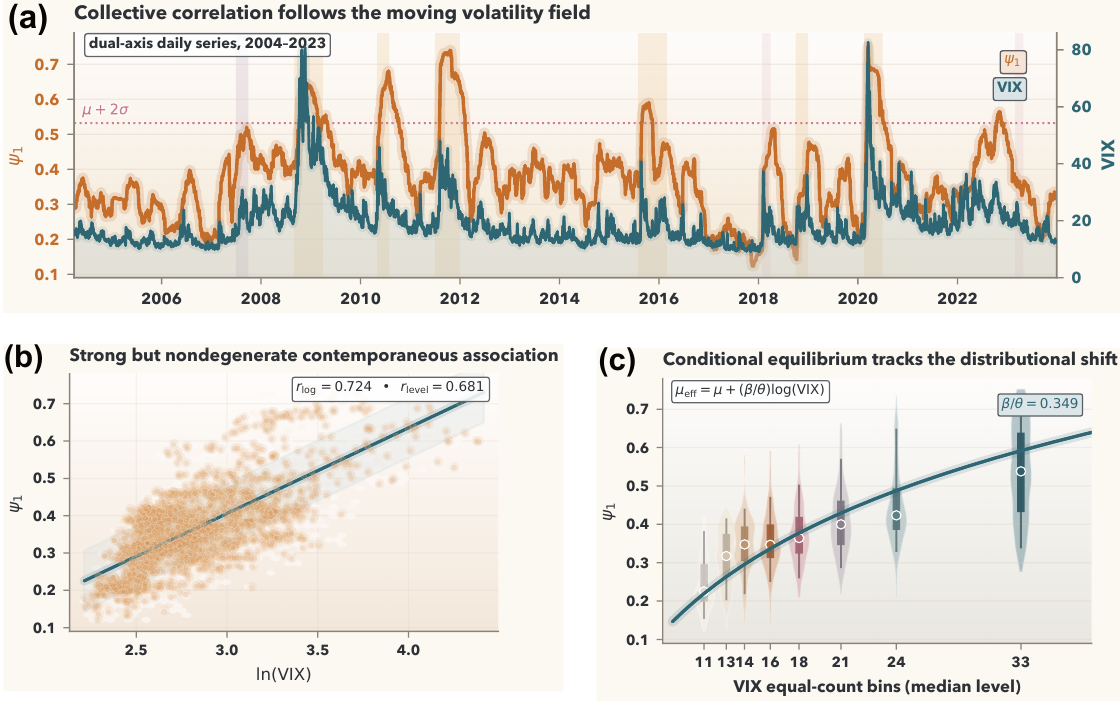}
\caption{\textbf{Collective correlation tracks a moving volatility field.}
(a) The long-run $\psi_1$ series co-moves strongly with VIX over 2004--2023,
with the largest excursions concentrated in major crisis episodes. (b) The
contemporaneous association is strong but nondegenerate, with
$r=0.724$ against $\log(\mathrm{VIX})$ and $r=0.681$ against level VIX.
(c) Equal-count VIX bins show that the fitted conditional equilibrium
$\mu_{\mathrm{eff}}(\mathrm{VIX})=\mu+(\beta/\theta)\log(\mathrm{VIX})$
tracks the empirical shift of the $\psi_1$ distribution across volatility
conditions. Together these panels motivate the attribution problem addressed
below.}
\label{fig:target}
\end{figure*}

The quantity to explain is not merely the existence of crisis episodes with
elevated collective correlation, but the fact that the full-sample
autocorrelation of $\psi_1$ remains elevated for months. Figure~\ref{fig:target}
shows that $\psi_1$ co-moves strongly with VIX across the 2004--2023 sample
while still exhibiting appreciable residual scatter around the observed
volatility field proxy. The question throughout is whether this slow-looking
behavior is intrinsic to $\psi_1$ or inherited from the motion of that
observed proxy.

\subsection{Conditional field-proxy decomposition}

The one-dimensional stochastic hierarchy yields the clearest result of the
paper. Table~\ref{tab:oned} summarizes the main model comparison. The
main reference model remains the VIX-coupled Ornstein--Uhlenbeck model
\cite{Risken1996,Gardiner2009,Cont2001},
\begin{equation}
d\psi_1=\left[-\theta(\psi_1-\mu)+\beta\log(\mathrm{VIX})\right]dt+\sigma\,dW,
\end{equation}
with best-fit parameters $\theta=0.01640$, $\mu=-0.6256$, $\beta=0.00572$,
and $\sigma=0.00942$. The associated effective equilibrium is
\begin{equation}
\mu_{\mathrm{eff}}(\mathrm{VIX})
=\mu+\frac{\beta}{\theta}\log(\mathrm{VIX}),
\end{equation}
which evaluates numerically to
$\mu_{\mathrm{eff}}(\mathrm{VIX})=-0.626+0.349\log(\mathrm{VIX})$ on the
aligned sample.
This linear approximation is intended only over the observed VIX range of
roughly $10$--$80$. Extrapolation far below that range would imply unphysical
negative equilibria once VIX falls below about $6$, which simply reflects the
local nature of the linearization.

At the baseline 60-day correlation-window construction, two timescales
separate sharply. In the bare OU model, $\theta_0=0.00335$, giving an
apparent relaxation time of roughly $298$ trading days. In the VIX-coupled OU
model, the intrinsic relaxation time conditional on the observed VIX proxy is
roughly $61$ trading days. The ratio $298/61\approx 4.9$ means that, in
OU-rate terms, conditioning on the observed VIX proxy removes about $79.5\%$
of the apparent persistence. These timescales should be
understood as
model-implied relaxation rates under the fitted linear OU approximation,
consistent with the pseudo-true interpretation adopted throughout.

\begin{table}
\caption{\label{tab:oned}Core one-dimensional and hybrid stochastic models on
the aligned daily sample.}
\begin{ruledtabular}
\begin{tabular}{lcc}
Model & BIC & $\Delta$BIC vs M2 \\
\hline
M\_RS,c+VIX & -33011.9 & -765.4 \\
M\_RS,c & -32274.4 & -27.9 \\
M2' & -32255.3 & -8.7 \\
M2 & -32246.5 & 0.0 \\
M3 & -32230.0 & +16.5 \\
M0 & -32137.5 & +109.0 \\
M1 & -32121.5 & +125.0 \\
\end{tabular}
\end{ruledtabular}
\end{table}

The global BIC winner is the hybrid constrained regime-switching plus VIX
model. Its fitted parameters, however, make its role quite specific: the calm
state carries $98.7\%$ of the stationary probability, the stress state lasts
only about $1.1$ trading days on average, and the continuous VIX loading
remains large ($\beta=0.00781$). In other words, the hybrid model is best
read as a rare extreme-event channel layered on top of the same continuous
field effect isolated by M2 \cite{Hamilton1989,AngBekaert2002}. The nested
likelihood-ratio test against the regime-switching baseline is also decisive
($\chi^2=746$, $p\approx 0$), so the hybrid fit validates the coexistence of
discrete extremes and continuous field motion rather than replacing the M2
interpretation \cite{Hamilton1989,White1982}. Within the hybrid model, the
calm-state parameters imply an intrinsic timescale of roughly $43$ trading
days. The corresponding timescale ratio is therefore
$1-43/298\approx 0.86$, slightly above the M2-based estimate of $0.80$. The
qualitative conclusion is unchanged: most apparent persistence is
field-inherited under either the M2 lens or the BIC-winning hybrid model.
Complete parameter estimates for M\_RS,c+VIX are reported in
Appendix Table~\ref{tab:rsvix_params_appendix}.

Two negative results are equally central. A quartic model without an explicit
field is strongly disfavored relative to the VIX-coupled OU model, and adding
quartic structure on top of VIX does not improve the fit. Within the tested
model family, the collective dynamics do not support a double-well or
Ising-like interpretation \cite{Bak1987,Sethna2001,Sornette2003}.

The state-dependent-noise variant M2$'$ provides only a modest refinement.
Allowing $\sigma(\psi_1)=\sigma_0+\sigma_1\psi_1$ improves BIC by $8.7$ units
relative to M2, indicating some structured heteroskedasticity, but it does
not improve the most difficult high-VIX PIT diagnostic. We therefore retain
the constant-$\sigma$ M2 model as the primary interpretive lens and read M2$'$
as a secondary adequacy check rather than as a change in the central
attribution story \cite{White1982,Gardiner2009}.

\begin{figure*}
\centering
\figpanel{0.98}{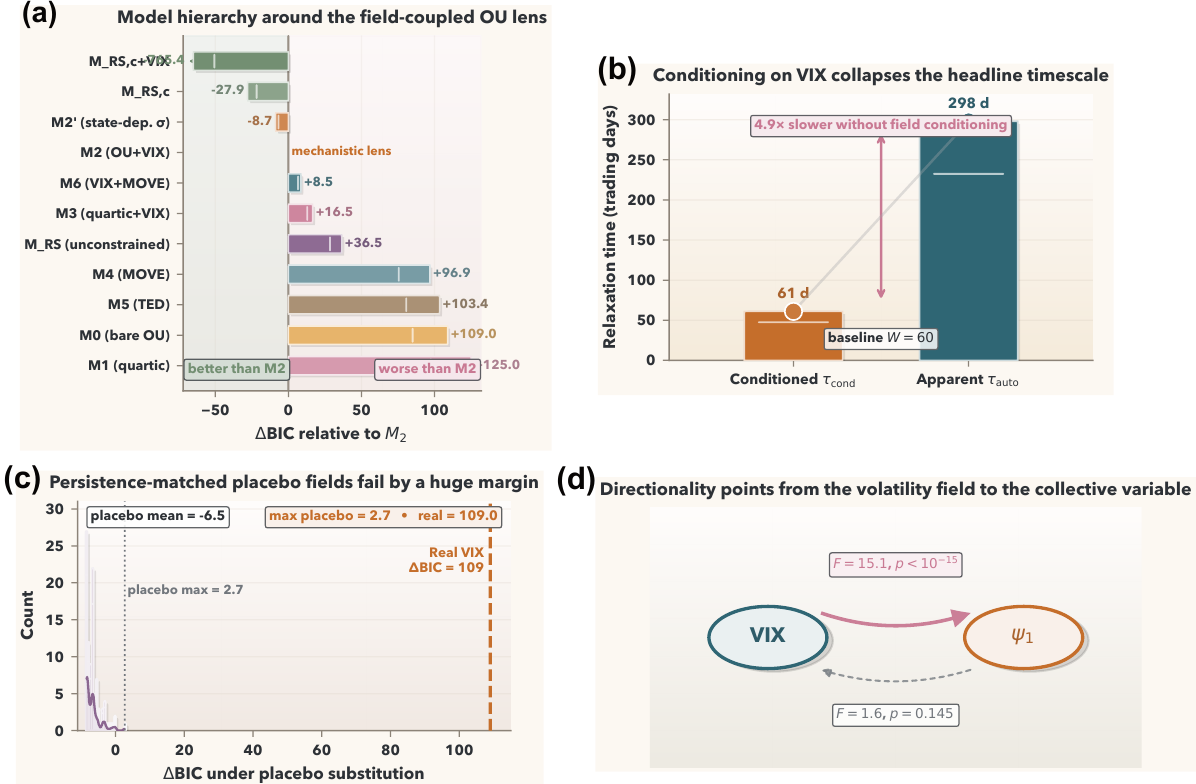}
\caption{\textbf{The main attribution result.} (a) Same-sample model
comparison across the core, hybrid, and auxiliary-field variants centers the
main interpretive story on the VIX-coupled OU model: bare OU and quartic
alternatives are strongly disfavored, while rare-event hybrids are better
read as an added extreme-state channel than as a replacement for the
continuous field picture. (b) Under the baseline $W=60$ construction,
conditioning on VIX reduces the relaxation time from $298$ to $61$ trading
days. (c) One hundred AR-matched placebo fields fail to approach the real-VIX
gain (maximum placebo $\Delta$BIC $=2.7$ versus $109$ for real VIX). (d)
Level-based Granger tests show dominant VIX$\rightarrow\psi_1$ directionality
($F=15.1$ versus $1.6$ in the reverse direction). These panels together
establish the conditional field-proxy attribution that anchors the rest of the
manuscript.}
\label{fig:attribution}
\end{figure*}

\subsection{Placebo falsification and robustness controls}

The strongest single discrimination test is the placebo-field experiment
\cite{Theiler1992}. We generate 100 AR(9)-matched synthetic fields that
reproduce the first-order persistence structure of $\log(\mathrm{VIX})$ and
compare their improvements against the real VIX field. None of the 100
surrogates matches the real VIX gain. The resulting distribution has mean
$\Delta$BIC$=-6.5$, standard deviation $2.19$, and maximum $2.70$, versus
$109.0$ for the real VIX field. This excludes an explanation based on
autocorrelation matching alone \cite{Theiler1992,White1982}.

Directionality supports the same picture. In the level-based Granger system,
VIX$\rightarrow\psi_1$ dominates with $F=15.07$ at lag 7, while the reverse
direction is not significant in levels. Under first differencing, VIX still
dominates ($F=10.63$ versus $2.76$), so the directional asymmetry is not a
simple rolling-window overlap artifact \cite{TodaYamamoto1995}.

The endogeneity objection is nevertheless real
\cite{ForbesRigobon2002,Cont2001}. Using the actual rolling correlation
structure together with fixed median stock volatilities, we estimate that
roughly $77.3\%$ of the shared explanatory power in the VIX--$\psi_1$
relationship is mechanical and $22.7\%$ is informational, with a partial
residual correlation of about $0.60$. Thus VIX is not a clean exogenous proxy
in the strict physical sense. But the informational increment is substantial,
and the placebo failure shows that it is not replaceable by generic
persistence.

The strongest new evidence comes from fitting these two components as
standalone fields. On the restricted sample where the split is defined, an M2
fit using the informational residual of $\log(\mathrm{VIX})$ alone still
improves on the bare OU benchmark by $\Delta$BIC$=78.6$, whereas the
mechanical proxy alone does not improve on the benchmark at all
($\Delta$BIC$=-8.3$). The full VIX field on that same restricted sample yields
$\Delta$BIC$=108.6$ (Appendix Table~\ref{tab:field_components_appendix}). In
other words, the part of VIX that carries model-selection power is
overwhelmingly informational rather than mechanical. The component gains are
not expected to add up exactly, since the OU parameters are re-estimated for
each effective field, but the dynamic implication is unambiguous: the
main field-proxy attribution is not being generated by trivial overlap
between realized volatility and realized correlation
\cite{ForbesRigobon2002,White1982}.

That dynamic conclusion is not tied to a single decomposition recipe. Across
three volatility-freezing choices (full-sample median, full-sample mean, and
pre-2016 median) and three weighting schemes (equal, inverse-volatility, and
volatility-share), the static mechanical fraction moves from $72.1\%$ to
$81.5\%$ while the informational fraction moves from $18.5\%$ to $27.9\%$
(Appendix Table~\ref{tab:decomp_sensitivity_appendix}). But the dynamic result
is invariant: informational-only fields remain strongly positive with
$\Delta$BIC in the narrow range $77.7$--$79.6$, whereas mechanical-only fields
remain BIC-negative in every tested variant
($-8.38\leq\Delta\mathrm{BIC}\leq -8.27$). The static split is therefore
operational rather than unique, while the dynamic conclusion is highly
stable across reasonable constructions \cite{White1982,ForbesRigobon2002}.

The overlapping-window concern can also be brought directly to the core
one-dimensional claim. Rebuilding the observable from disjoint 5-trading-day
correlation matrices sampled every 5 trading days, the field-coupled OU still
beats bare mean reversion by $\Delta$BIC$=151.2$ for weekly $\psi_1$ and by
$\Delta$BIC$=140.1$ for weekly mean market correlation
(Appendix Table~\ref{tab:nonoverlap_appendix}). On those disjoint weekly
series the effective timescales shorten, as expected from the coarser clock,
but the sign of the attribution result is unchanged: SCPA remains positive at
$0.286$ and $0.282$, respectively. At the opposite extreme, a same-horizon
disjoint 60-day block reconstruction yields only $\Delta$BIC$=1.7$ when paired
with end-of-block VIX and $\Delta$BIC$=-3.4$ when paired with block-mean VIX.
We therefore interpret that 60-day block version as a low-power boundary check
with only 77 observations, not as a power-matched replacement for the daily
series. The main point is narrower but decisive: once overlap is removed in a
reasonably information-preserving weekly reconstruction, the field-coupled
model remains strongly preferred
\cite{Wand2023Entropy,Wand2023Arxiv,White1982}.

Cross-market auxiliary fields are weaker. MOVE provides only a modest
same-sample gain ($\Delta$BIC$=12.0$ versus the bare model), while TED does
not improve the fit ($\Delta$BIC$=-1.1$). Several reasonable transformations
of MOVE yield the same conclusion: it is auxiliary, not decisive. The
strongest cross-field statement the current evidence supports is therefore
modest: VIX is the uniquely strong field proxy in this dataset, MOVE is
secondary, and TED does not help.

The stochastic fits are complemented by two model-free controls. The first is
a quiet-regime autocorrelation check. The physically motivated strict
criterion---daily VIX inside $[15,18]$ for at least 120 consecutive trading
days---yields no qualifying segment in the entire sample. That absence is
itself informative: the volatility field is never truly dormant for long,
consistent with continual modulation of $\psi_1$
\cite{Scheffer2009,Cont2001}. To obtain a finite-sample quiet-regime
estimate, we therefore adopt a transparent relaxed criterion based on a
20-day rolling-median VIX in $[14,20]$, which yields three qualifying
segments. Under that definition the e-folding lag drops from $69$ days to
$27$ days, with an episode-bootstrap $95\%$ confidence interval of
$[18,33]$ days, the ACF at lag 20 falls from $0.853$ to $0.514$, and by lag
40 the quiet-regime ACF has already crossed below zero. The same $27$-day e-folding
lag is recovered under the wider $[13,21]$ band, whereas the narrower
$[15,19]$ band yields no qualifying 120-day segment at all
(Appendix Table~\ref{tab:quiet_bands_appendix}). The quiet-regime diagnostic
is therefore partially robust: it is replicated across the two widest tested
bands, but finite-sample support disappears at the narrowest band.

The second control is a field-stripped residual. Defining
\begin{equation}
\epsilon_{\mathrm{M2}}(t)=\psi_1(t)-\mu_{\mathrm{eff}}(\mathrm{VIX}(t)),
\end{equation}
we find that the residual remains persistent, but much less so than raw
$\psi_1$: the e-folding lag drops from $69$ to $42$ days, the integrated ACF
up to lag 60 drops from $44.3$ to $30.6$, and the integrated ACF up to lag 90
drops from $54.2$ to $38.7$. Field stripping therefore does not annihilate
all persistence, but it removes a substantial fraction of it.

These controls are supported by a broader out-of-sample check. The baseline
split at 2016-01-01 leaves the per-observation log likelihood of M2 almost
unchanged between train and test periods ($3.252$ versus $3.235$, ratio
$0.995$), which argues against the main result being a crisis-specific
in-sample artifact. Because the post-2016 holdout contains the COVID-19 spike,
we also repeat the evaluation after excluding 2020-03-01 through 2020-09-30
and scoring the two remaining contiguous test segments separately. The test log
likelihood per observation then rises to $3.310$, with a test/train ratio of
$1.018$, so the stability claim does not depend on the COVID episode. More
importantly, a six-way anchored split sweep at 2010, 2012, 2014, 2016, 2018,
and 2020 leaves the sign of the result unchanged on every held-out remainder:
M2 beats M0 on test in all six cases, with per-observation test-set gains
ranging from $0.0023$ to $0.0152$ and M2 test/train ratios ranging from
$0.964$ to $0.999$ (Appendix Table~\ref{tab:oos_splits_appendix}). The
field-coupled advantage is therefore not tied to a single historical partition
\cite{White1982}.

An exact raw-return window sweep sharpens the interpretation of the headline
timescale. Recomputing $\psi_1$ directly from returns for
$W=30,45,60,90,120$ leaves the field advantage intact at every window
($\Delta$BIC$=99$--$165$ for M2 over M0) and keeps the field susceptibility
$\beta/\theta$ in a narrow range of $0.336$--$0.376$
(Appendix Table~\ref{tab:window_robustness_appendix}). The absolute
conditioned timescale is not window-invariant: it rises from $34$ trading days
at $W=30$ to $97$ trading days at $W=120$, while the bare timescale rises from
$105$ to $839$ days. This behavior is expected from the low-pass filtering
inherent in rolling-window estimation. The more structural quantity is the
susceptibility $\beta/\theta$, which determines the equilibrium displacement
per unit of $\log(\mathrm{VIX})$ and remains stable across all tested windows.
The attribution fraction itself, measured by
$1-\tau_{\mathrm{cond}}/\tau_{\mathrm{auto}}$, remains majority-inherited at
every window but increases from $0.67$ to $0.88$ rather than staying strictly
constant. The $61$-day figure should therefore be read as the effective
conditioned timescale at the baseline resolution $W=60$, not as a
window-independent physical constant \cite{Gardiner2009,White1982}. The
monotonic increase of SCPA with $W$ is itself consistent with rolling-window
low-pass filtering, which inflates the bare timescale more strongly because
the bare model must absorb field-driven slow motion that the conditioned model
removes explicitly. The formal limit $W\rightarrow 0$ is not an operational
object for rolling-correlation estimation: while the leading eigenvalue
fraction remains estimable in the high-dimensional $W<N$ regime, arbitrarily
short windows would collapse the statistic into finite-sample noise
\cite{Bun2017}. The smallest analyzed window here is $W=30$, where the bare
timescale is still $105$ trading days, already well above the window length
itself.

\begin{figure*}
\centering
\figpanel{0.98}{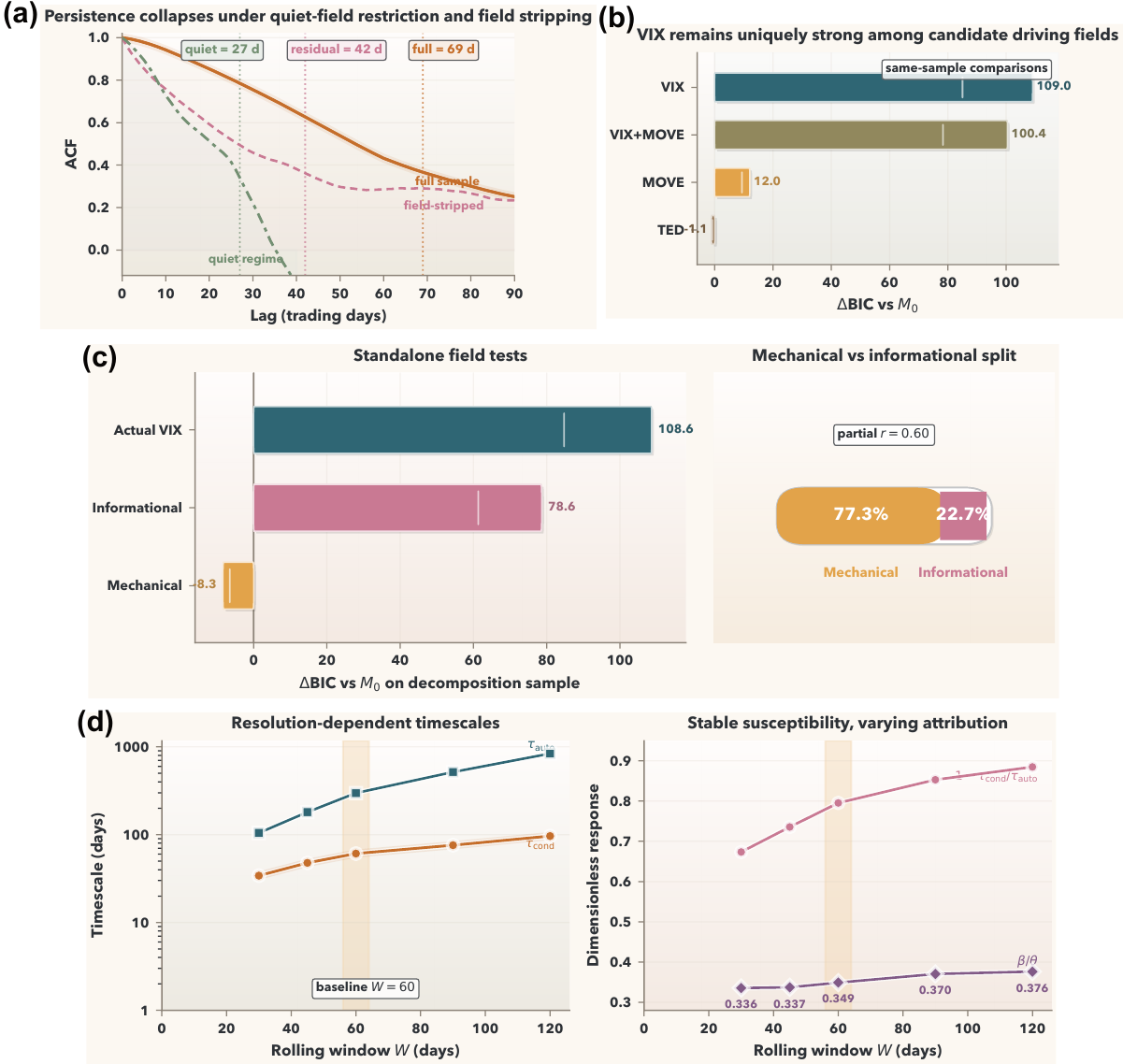}
\caption{\textbf{Independent robustness checks.} (a) Persistence collapses
both in quiet regimes and after subtracting the fitted VIX-dependent
equilibrium: the full-sample e-folding lag is $69$ trading days, the
field-stripped residual falls to $42$ days, and the quiet-regime estimate
falls to $27$ days. (b) Among auxiliary fields, VIX remains uniquely strong,
while MOVE is secondary and TED does not improve on the bare benchmark. (c)
On the decomposition sample, the informational residual of
$\log(\mathrm{VIX})$ alone retains most of the model-selection gain
($\Delta$BIC $=78.6$), whereas the mechanical proxy alone fails to improve
the fit ($\Delta$BIC $=-8.3$); the same sample still yields a
$77.3\%/22.7\%$ mechanical/informational variance split. (d) Exact
recomputation across $W=30$--$120$ shows that absolute timescales depend on
rolling-window resolution, but the field susceptibility $\beta/\theta$ stays
in the narrow range $0.336$--$0.376$ and the attribution fraction remains
majority-inherited at every window. Together these panels show that the main
claim is not an artifact of placebo persistence, auxiliary fields, mechanical
overlap, or the baseline window choice.}
\label{fig:robustness}
\end{figure*}

\subsection{Two-dimensional hidden-variable extension}

The strongest theoretical extension was to ask whether the observed VIX can
itself serve as the hidden variable behind one-dimensional generalized
Langevin memory \cite{Wand2023Entropy,Wand2023Arxiv,Zwanzig2001}. The
following extension is exact only within the linear-Gaussian class, with
nonlinear departures treated as a limitation rather than silently absorbed
into the interpretation \cite{Gardiner2009,White1982}. To test this, we fit
an exact linear-Gaussian two-dimensional system,
\begin{align}
d\psi_1 &= \left[-\theta_{\psi}(\psi_1-\mu_{\psi})+\beta_{\psi}v\right]dt
 + \sigma_{\psi}dW_1,\\
dv &= \left[-\theta_v(v-\mu_v)+\beta_v\psi_1\right]dt + \sigma_v dW_2,
\end{align}
with $v=\log(\mathrm{VIX})$ and correlated innovations. Because the one-step
transition density is exactly Gaussian, the discrete VAR(1) likelihood is the
exact transition-density MLE within this model family.

On the aligned daily series, feedforward coupling is preferred over
bidirectional drift by $\Delta$BIC$=2.4$, while the feedforward model beats
the decoupled model by $\Delta$BIC$=125.3$. The $\Delta$BIC$=2.4$ margin
between feedforward and bidirectional drift is small by conventional
model-selection standards \cite{KassRaftery1995}, so the daily data offer
only a weak preference for the simpler feedforward form. The preferred daily
model reproduces the one-dimensional M2 parameters on the $\psi_1$ side almost
exactly, but sets $\beta_v=0$ in the best structure. On the daily sample,
observed VIX behaves as a persistent driver, not as a preferred hidden
variable whose elimination generates a self-memory kernel
\cite{Zwanzig2001,Wand2023Entropy}.

The key stress test is therefore the Wand-faithful weekly construction. The
daily series is built from overlapping windows, which can accentuate serial
dependence and may sharpen the appearance of one-way drive. We therefore
rebuild the problem using disjoint 5-day correlation windows sampled every
5 trading days, much closer to the preprocessing in the generalized-Langevin
literature \cite{Wand2023Entropy,Wand2023Arxiv}. Under this construction the
evidence changes: a weekly $\psi_1$ proxy weakly prefers the full
bidirectional model by $\Delta$BIC$=0.78$ over feedforward, and the weekly
mean market correlation weakly prefers it by $\Delta$BIC$=2.69$.

This is the strongest surviving part of the hidden-variable story:
bidirectional structure becomes more plausible once the comparison is made
under a preprocessing regime close to the prior weekly work. But the bridge
still stops short of full identification \cite{Zwanzig2001,Falkena2019}.
Solving the second equation formally and substituting into the first yields a
projected memory term of the form
\begin{equation}
K(t)=\beta_{\psi}\beta_v e^{-\theta_v t},
\end{equation}
plus colored-noise contributions. The implied timescales are not close to the
conservative $\sim 15$-day benchmark implied by ``three trading weeks'' in
Wand \textit{et al.} On our outputs they are approximately $36.5$ days for
the daily full model, $9.1$ days for naive weekly thinning, $33.5$ days for
the Wand-faithful weekly $\psi_1$ proxy, and $36.5$ days for the Wand-faithful
mean market correlation. The bridge is therefore nontrivial but incomplete: a
more faithful weekly comparison makes bidirectional coupling more plausible,
yet observed VIX still does not recover the reported generalized-Langevin
memory scale.

The 2D placebo gate still adds independent value. The daily feedforward model
beats the daily decoupled model by roughly $125$ BIC units, and none of 100
AR-matched placebo fields reproduces the real coupling gain. So the 2D
section remains essential even though it supports only partial reconciliation
rather than full unification.

\begin{figure*}
\centering
\figpanel{0.98}{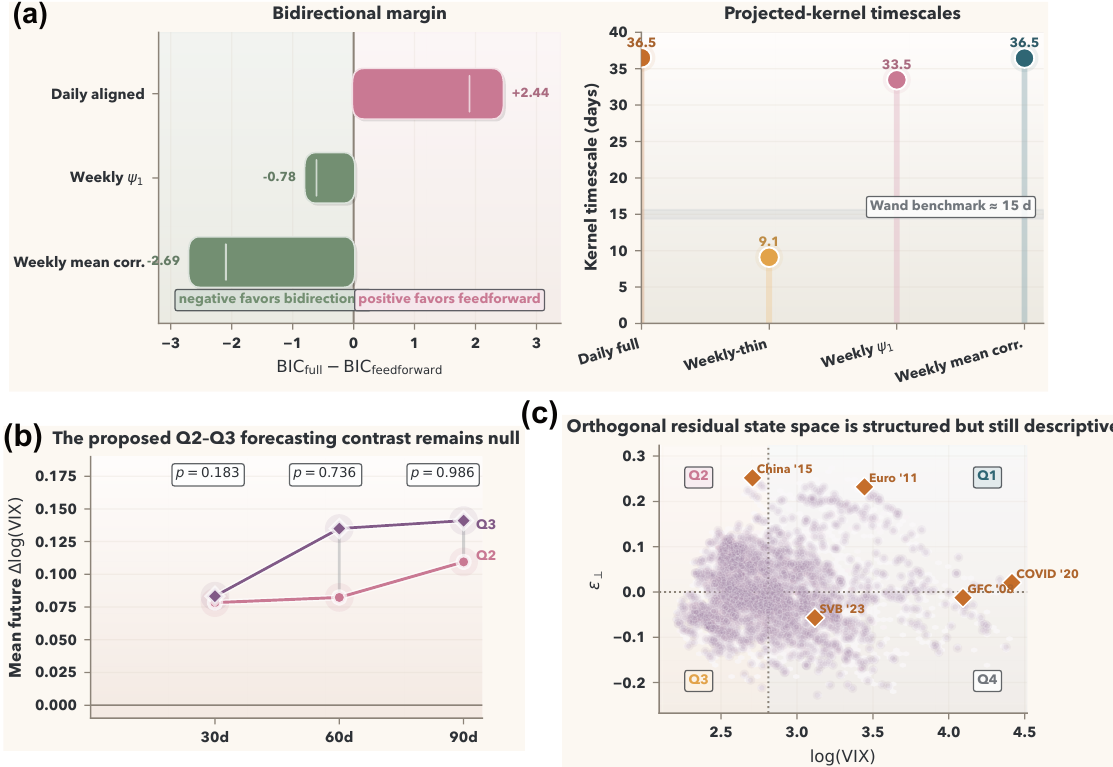}
\caption{\textbf{Extensions beyond the main one-dimensional result.} (a) On
daily aligned data the 2D comparison weakly favors feedforward over
bidirectional drift by $\Delta$BIC $=2.44$, whereas the Wand-faithful weekly
reconstructions weakly favor bidirectionality ($\Delta$BIC $=-0.78$ for
weekly $\psi_1$ and $-2.69$ for weekly mean correlation; negative values
favor bidirectional coupling). The projected-kernel timescales remain in the
$9$--$36$ day range and therefore do not recover the $15$-day Wand benchmark
in any clean way. (b) The orthogonal residual phase space in
$(\log\mathrm{VIX},\epsilon_\perp)$ is structured and visually nontrivial.
(c) The proposed Q2-versus-Q3 prediction test nevertheless remains null
across all tested horizons ($p=0.183$, $0.736$, and $0.986$ at 30, 60, and
90 trading days). The figure therefore frames these sections as constrained
extensions rather than as new headline results.}
\label{fig:extensions}
\end{figure*}

\subsection{Orthogonal residual analysis}

The final analytical step asks whether a residual order parameter beyond VIX
can carry operational information. Two residuals are useful to distinguish
\cite{White1982,Driessen2009}. The mechanistic residual,
\begin{equation}
\epsilon_{\mathrm{M2}}(t)=\psi_1(t)-\mu_{\mathrm{eff}}(\mathrm{VIX}(t)),
\end{equation}
is natural for field-stripping but not orthogonal to VIX in levels. The
relevant candidate order parameter is instead the orthogonal level residual,
\begin{equation}
\epsilon_{\perp}(t)=\psi_1(t)-\left(a+b\log(\mathrm{VIX}(t))\right),
\end{equation}
where $(a,b)$ are estimated from a direct level regression and therefore
enforce $\mathrm{corr}(\epsilon_{\perp},\log \mathrm{VIX})\approx 0$.

This variable is not trivial noise. Its e-folding lag is about $43$ days,
compared with $69$ days for raw $\psi_1$, and its integrated ACF to lag 60 is
$32.1$ instead of $44.3$. In that sense, $\epsilon_{\perp}$ is a genuine
residual mode rather than a negligible leftover.

The associated phase-space program is only partially successful. Partitioning
the $(\log\mathrm{VIX},\epsilon_{\perp})$ plane by the sample median of
$\log\mathrm{VIX}$ and the sign of $\epsilon_{\perp}$ yields balanced time
fractions overall, but event-peak classification is less rich than hoped: Q1
contains six peak events, Q4 contains three, Q2 contains only China/Oil
2015--16, and Q3 contains none of the predefined event peaks. Thus the phase
space is descriptive, but not a mature four-state taxonomy.

The strongest proposed operational test also fails. Comparing
low-VIX/high-$\epsilon_{\perp}$ (Q2) days against low-VIX/low-$\epsilon_{\perp}$
(Q3) days, we find no evidence that Q2 precedes larger future VIX increases at
30, 60, or 90 trading days. The mean differences are small, the
Mann--Whitney tests are not significant, and the associated rank-biserial
effect sizes remain close to zero (Table~\ref{tab:q2q3}). The orthogonal
residual therefore survives as a descriptive state variable, but not as a
demonstrated forecasting signal.

\begin{table}
\caption{\label{tab:q2q3}Future VIX changes after low-VIX days split by the
sign of the orthogonal residual. Positive rank-biserial values would favor the
proposed Q2 signal.}
\begin{ruledtabular}
\begin{tabular}{ccccc}
Horizon & Mean Q2 & Mean Q3 & Mann--Whitney $p$ & Rank-biserial \\
\hline
30 days & 0.078 & 0.083 & 0.183 & 0.021 \\
60 days & 0.082 & 0.135 & 0.736 & -0.015 \\
90 days & 0.110 & 0.141 & 0.986 & -0.052 \\
\end{tabular}
\end{ruledtabular}
\end{table}

Taken together, the results define a deliberately uneven evidence profile: the
conditional field-proxy decomposition is strongly supported, the two-dimensional bridge
is only partially supported, and the orthogonal residual is descriptive
rather than predictive.

\section{Discussion}

The central result is conditional rather than fully causal: within the tested
stochastic class, most apparent persistence in market collective correlation is
absorbed once one conditions on an observed VIX field proxy rather than
leaving the dynamics bare. The strongest contribution of the paper is
therefore not a universal hidden-variable model, but a falsifiable attribution
protocol that sharply narrows the interpretation of slow collective dynamics
\cite{White1982,Theiler1992,Zwanzig2001}. A field-coupled OU model, a placebo
gate that rejects autocorrelation-matched surrogates, disjoint weekly
reconstructions that preserve the M2 advantage, decomposition variants that
leave informational-only fields strongly positive while mechanical-only fields
remain negative, and both quiet-regime and field-stripped ACF collapse all
point in the same direction. This convergence is unusually strong for a
soft-system inference problem
\cite{ForbesRigobon2002,Bekaert2005,LeBaron2001,Bloom2009}.

\subsection{Reconciling persistence attribution measures}

Two quantitative attribution measures appear in this paper and should not be
conflated. The timescale ratio
$1-\tau_{\mathrm{cond}}/\tau_{\mathrm{auto}}\approx 0.80$ compares
model-implied OU relaxation rates before and after conditioning on VIX
\cite{Risken1996,Gardiner2009}. By contrast, the field-stripped residual ACF
reduction of about $39\%$ compares the e-folding lag of raw $\psi_1$ against
that of $\epsilon_{\mathrm{M2}}=\psi_1-\mu_{\mathrm{eff}}(\mathrm{VIX})$,
dropping from $69$ to $42$ trading days. The first quantity is a model-based
rate comparison within the OU class; the second is a model-free comparison of
the temporal structure left after subtracting the fitted conditional
equilibrium.

These numbers need not agree. The timescale ratio measures how fast $\psi_1$
relaxes toward a moving VIX-dependent target once that target is specified.
The residual ACF measures how much persistence remains after that target is
subtracted, including persistence from imperfect linearization, additional
latent drivers, and rolling-window measurement structure
\cite{White1982,Gardiner2009}. They therefore bracket the field effect from
different angles rather than estimating a single universal percentage.

A further caveat is that the absolute OU timescales are resolution-dependent.
Exact recomputation with $W=30$--$120$ shows that the conditioned relaxation
time increases from $34$ to $97$ trading days as the rolling window
lengthens, which is consistent with the low-pass filtering built into
rolling-window observables \cite{Gardiner2009,White1982}. The more structural
quantity is the susceptibility $\beta/\theta$, which remains stable across
those windows in the narrow range $0.336$--$0.376$. By contrast, the
model-based attribution fraction $1-\tau_{\mathrm{cond}}/\tau_{\mathrm{auto}}$
is not strictly invariant: it stays in the majority-inherited regime for
every window, but rises from about $0.67$ to $0.88$ as $W$ increases. The
baseline $61$-day estimate should therefore be read as the effective
conditioned timescale of the standard $W=60$ construction, not as a
window-free microscopic constant.

The two-dimensional extension refines that conclusion without overturning it.
On the aligned daily sample, observed VIX behaves as a persistent driver and
not as the preferred hidden coordinate whose projection would reproduce
generalized-Langevin memory. At the same time, the Wand-faithful weekly
reconstruction does revive weak bidirectional preference, though still within
the inconclusive BIC range, so the relation to Wand \textit{et al.} is more
nuanced than simple contradiction. We also note
that the Wand sample (1992--2012, $N=249$) differs from ours in both period
and stock universe, so part of the kernel-timescale discrepancy may reflect
data differences rather than a fundamental failure of the Mori--Zwanzig
identification. The most defensible synthesis is partial reconciliation:
explicit field driving and hidden-variable memory are not unrelated
descriptions, but observed VIX alone does not complete the Mori--Zwanzig
identification
\cite{Wand2023Entropy,Wand2023Arxiv,Zwanzig2001,Falkena2019}.

The orthogonal residual extension leads to a parallel lesson. Constructing
$\epsilon_{\perp}$ is easy, and its dynamics are not trivial. It yields a
genuine VIX-orthogonal residual mode with intermediate persistence and some
event-classification structure. But that is not the same as demonstrating an
operationally useful predictive order parameter \cite{White1982}. The
negative Q2-versus-Q3 tests show why it is worth keeping those claims
separate. This residual also differs conceptually from the implied-correlation
versus realized-correlation gaps studied in the correlation-risk-premium
literature \cite{Driessen2009}: here the object is an orthogonalized
collective state variable, not an option-pricing spread.

These results sharpen the relation to the criticality literature. The present
evidence does not support a double-well, Ising-like, or critical-slowing
interpretation of $\psi_1$ dynamics
\cite{Sornette2003,LuxMarchesi1999,ContBouchaud2000,Bouchaud2013,MarsiliRaffaelli2006,Scheffer2009}.
Within the tested stochastic class, the physical picture is closer to a
driven single-well system with persistent field motion and a modest
complementary extreme-state channel \cite{Risken1996,Gardiner2009}. Strong
visual persistence of the collective variable is therefore not, by itself,
evidence for a near-critical endogenous transition
\cite{Scheffer2009,White1982}.

This also clarifies the relation to earlier Kramers--Moyal extractions for
financial correlation observables. Unlike Stepanov \textit{et al.}, who
extracted Kramers--Moyal coefficients for a related dominating correlation
variable without testing external fields explicitly, the present work
conditions on candidate drivers and subjects them to placebo-surrogate
discrimination \cite{Stepanov2015,Theiler1992}.

More broadly, the methodology should transfer to other soft complex systems.
Dynamic functional connectivity in fMRI or EEG, delayed climate indices, and
ecological synchrony measures all face the same question: is the slow
observable intrinsically slow, or is it being dragged by a persistent field
proxy
\cite{Allen2014,Scheffer2009,Falkena2019}? The same placebo-controlled
sequence used here can be applied there as well: compare bare and
field-coupled stochastic models, reject persistence-only surrogates,
quantify mechanical overlap where needed, and only then test stronger
hidden-variable or residual-state interpretations
\cite{Theiler1992,Zwanzig2001,Falkena2019}. Exact multivariate OU fits also
open a route to nonequilibrium diagnostics such as entropy-production measures
\cite{Gilson2023}.

Abstracted away from finance, the protocol has four generic ingredients: a
collective observable, an observed candidate field proxy on the same clock, a
field-free versus field-coupled stochastic comparison with matched likelihood
treatment, and an adversarial surrogate or decomposition stage that attacks
persistence-only and mechanical-only explanations
\cite{Theiler1992,White1982,Zwanzig2001}. It is strongest when the observable
can be reconstructed at multiple resolutions and when a non-overlapping
version is available. Its generic failure mode is latent common drive: if an
unobserved $z$ moves both the proxy and the observable, the protocol can still
show that conditioning on the proxy removes apparent persistence, but it
cannot by itself prove direct coupling from proxy to observable
\cite{White1982,Zwanzig2001,Falkena2019}.

\subsection{Limitations}

Several limits remain important. First, the VIX relationship is materially
mechanical even after the decomposition \cite{ForbesRigobon2002,Cont2001}.
The static mechanical share is operational rather than unique: across
reasonable freeze and weighting schemes it ranges from $0.72$ to $0.81$, even
though the dynamic model-selection conclusion remains invariant.
Second, MOVE and TED do not support a broader cross-market hierarchy beyond
the main VIX result. Third, the two-dimensional extension is exact only within
a linear-Gaussian class, and the state-dependent-noise diagnostics indicate
structured inadequacies in extreme regimes \cite{Gardiner2009,White1982}.
Fourth, the event sample is still small relative to the breadth of financial
stress phenomenology. Finally, the main failure mode of the attribution
protocol is a latent common driver that is only imperfectly proxied by the
observed field proxy; placebo rejection against the observed proxy does not
eliminate that possibility \cite{White1982,Zwanzig2001}. More precisely, the
placebo gate tests whether the observed proxy carries coupling information
beyond marginal persistence, but it does not by itself distinguish direct
coupling $u\rightarrow y$ from a confounded structure in which an unobserved
driver $z$ influences both the proxy $u$ and the observable $y$.

These caveats define the paper's boundary rather than its weakness. The
manuscript is strongest on conditional field-proxy attribution, more limited on the
hidden-variable bridge, and explicitly negative on operational residual
prediction. These boundaries delineate what can and cannot be concluded from
the present evidence.

\section*{Data availability}

Daily VIX closes were obtained from Federal Reserve Economic Data. The
processed datasets needed to reproduce the figures and tables in this
manuscript are available from the authors on reasonable request.

\section*{Code availability}

The analysis and figure-generation code used for this manuscript are
available from the authors on reasonable request.

\clearpage
\onecolumngrid
\appendix

\section{Exact Likelihoods and Analytical Derivations}
\label{sec:app_exact}

\subsection{Exact one-dimensional transition laws}

For the bare OU model,
\begin{equation}
d\psi_t=-\theta_0(\psi_t-\mu_0)\,dt+\sigma_0\,dW_t,
\end{equation}
the exact solution over one unit daily interval is
\begin{equation}
\psi_{t+1}=e^{-\theta_0}\psi_t+\left(1-e^{-\theta_0}\right)\mu_0+\eta_t,
\end{equation}
with Gaussian innovation
\begin{align}
\eta_t &\sim \mathcal{N}\!\left(0,q_0\right),\\
q_0 &= \frac{\sigma_0^2}{2\theta_0}\left(1-e^{-2\theta_0}\right),
\end{align}
which is the standard exact transition density of the Ornstein--Uhlenbeck
process \cite{Risken1996,Gardiner2009}. The one-step log likelihood is
\begin{align}
\ell_{\mathrm{M0}}
&= -\frac{n-1}{2}\log(2\pi q_0) \nonumber\\
&\quad - \frac{1}{2q_0}\sum_{t=1}^{n-1}
\left(\psi_{t+1}-m^{(0)}_t\right)^2,
\end{align}
where $m^{(0)}_t=e^{-\theta_0}\psi_t+(1-e^{-\theta_0})\mu_0$.

For the field-coupled model,
\begin{equation}
d\psi_t=\left[-\theta(\psi_t-\mu)+\beta v_t\right]dt+\sigma\,dW_t,
\end{equation}
we treat the observed field value $v_t=\log(\mathrm{VIX}_t)$ as constant on
the interval $[t,t+1)$. Variation of constants gives
\begin{multline}
\psi_{t+1}=e^{-\theta}\psi_t \\
+ \int_t^{t+1}e^{-\theta(t+1-s)}
\left(\theta\mu+\beta v_t\right)\,ds
+ \int_t^{t+1}e^{-\theta(t+1-s)}\sigma\,dW_s.
\end{multline}
The deterministic integral evaluates to
\begin{equation}
\left(1-e^{-\theta}\right)
\left(\mu+\frac{\beta}{\theta}v_t\right),
\end{equation}
and the stochastic integral is Gaussian with variance
$\sigma^2(1-e^{-2\theta})/(2\theta)$ \cite{Risken1996,Gardiner2009}. Hence
\begin{align}
\psi_{t+1}\mid \psi_t,v_t &\sim \mathcal{N}\!\left(m^{(2)}_t,q\right),\\
m^{(2)}_t
&= e^{-\theta}\psi_t
+\left(1-e^{-\theta}\right)
\left(\mu+\frac{\beta}{\theta}v_t\right),\\
q &= \frac{\sigma^2}{2\theta}\left(1-e^{-2\theta}\right).
\end{align}
The corresponding log likelihood is
\begin{align}
\ell_{\mathrm{M2}}
&= -\frac{n-1}{2}\log(2\pi q) \nonumber\\
&\quad - \frac{1}{2q}\sum_{t=1}^{n-1}
\left(\psi_{t+1}-m^{(2)}_t\right)^2.
\end{align}
The quantities reported in the main text are then
\begin{align}
\tau_{\mathrm{auto}} &= \frac{1}{\theta_0},\\
\tau_{\mathrm{cond}} &= \frac{1}{\theta},\\
\chi &= \frac{\beta}{\theta},\\
\mu_{\mathrm{eff}}(v) &= \mu+\chi v,
\end{align}
where $\chi$ is the field susceptibility, i.e. the equilibrium displacement
per unit of $\log(\mathrm{VIX})$.

\subsection{Mechanical and informational field decomposition}

Let $m_t$ denote the mechanical volatility proxy and
$a_t=\log(\mathrm{VIX}_t)$ denote the actual field. The informational residual
is defined by the auxiliary regression
\begin{equation}
a_t=\gamma_0+\gamma_1 m_t+r_t,
\end{equation}
so that $r_t$ is orthogonal to the fitted mechanical component in sample. We
then fit the sequential regressions
\begin{align}
\psi_t &= \alpha_0+\alpha_1 m_t+u_t,\\
\psi_t &= \beta_0+\beta_1 m_t+\beta_2 r_t+\varepsilon_t.
\end{align}
If $R^2_{\mathrm{mech}}$ is the coefficient of determination of the first
regression and $R^2_{\mathrm{full}}$ that of the second, the reported
fractions are
\begin{align}
f_{\mathrm{mech}} &= \frac{R^2_{\mathrm{mech}}}{R^2_{\mathrm{full}}},\\
f_{\mathrm{info}}
&= \frac{R^2_{\mathrm{full}}-R^2_{\mathrm{mech}}}{R^2_{\mathrm{full}}}.
\end{align}
With the decomposition sample used in the manuscript,
$R^2_{\mathrm{full}}=0.7120$, $R^2_{\mathrm{mech}}=0.5505$, and the
informational increment is $0.1615$, yielding
$f_{\mathrm{mech}}=0.7732$ and $f_{\mathrm{info}}=0.2268$. These quantities
summarize shared explanatory power in a static regression sense; they do not
imply that the standalone OU-field improvements must add linearly, because
the M2 parameters are re-estimated for each field and the fitted equilibrium
map changes with the field definition \cite{White1982}. That is precisely why
the separate M2 fits with actual, mechanical-only, and informational-only
fields are reported as distinct experiments rather than algebraically
combined.

\subsection{Two-dimensional OU projection and induced memory kernel}

Write the centered two-dimensional system as
\begin{align}
\dot{\mathbf{x}}(t) &= A\mathbf{x}(t)+\Sigma \boldsymbol{\xi}(t),\\
\mathbf{x}(t)
&=
\begin{pmatrix}
\psi_1(t)-\mu_{\psi}\\
v(t)-\mu_v
\end{pmatrix},
\end{align}
with drift matrix
\begin{equation}
A=
\begin{pmatrix}
-\theta_{\psi} & \beta_{\psi}\\
\beta_v & -\theta_v
\end{pmatrix}.
\end{equation}
The exact discrete transition over one interval $\Delta$ is
\begin{align}
\mathbf{x}_{t+\Delta}
&=e^{A\Delta}\mathbf{x}_t+\boldsymbol{\eta}_t,\\
\boldsymbol{\eta}_t
&\sim \mathcal{N}\!\left(0,Q_{\Delta}\right),
\end{align}
with covariance
\begin{equation}
Q_{\Delta}=\int_0^{\Delta}
e^{As}\Sigma\Sigma^{\top}e^{A^{\top}s}\,ds,
\end{equation}
which is the exact continuous-time/discrete-time correspondence for
linear-Gaussian systems \cite{Gardiner2009,Zwanzig2001}. The structural
restrictions used in the paper are: decoupled
($\beta_{\psi}=\beta_v=0$), feedforward
($\beta_{\psi}\neq 0,\beta_v=0$), and bidirectional
($\beta_{\psi}\neq 0,\beta_v\neq 0$). The daily and weekly comparisons are
carried out by exact Gaussian likelihood in this discrete representation, and
the continuous-time coefficients are obtained by matrix logarithm of the
estimated transition matrix.

To expose the projected-memory interpretation, solve the second equation
formally:
\begin{multline}
v(t)=v(0)e^{-\theta_v t} \\
+ \int_0^t e^{-\theta_v(t-s)}
\left[\beta_v\psi_1(s)\,ds+\sigma_v\,dW_2(s)\right].
\end{multline}
Substituting this expression into the observable equation yields
\begin{align}
\dot{\psi}_1(t)
={}& -\theta_{\psi}\psi_1(t)+\beta_{\psi}v(0)e^{-\theta_v t}\nonumber\\
&+\beta_{\psi}\beta_v\int_0^t e^{-\theta_v(t-s)}\psi_1(s)\,ds \nonumber\\
&+\beta_{\psi}\sigma_v\int_0^t e^{-\theta_v(t-s)}\,dW_2(s)
+\sigma_{\psi}\xi_1(t).
\end{align}
The induced self-memory kernel is therefore
\begin{equation}
K(t)=\beta_{\psi}\beta_v e^{-\theta_v t},
\end{equation}
and the associated projected timescale is $\tau_K=1/\theta_v$
\cite{Zwanzig2001,Falkena2019}. When $\beta_v=0$, as in the
daily-preferred feedforward structure, this kernel vanishes identically. The
observed VIX can only generate a nonzero projected memory kernel when the
bidirectional term is empirically supported.

\subsection{Definitions of attribution measures}

Two attribution measures are used repeatedly in the manuscript. The first is
the model-based scalar persistence-collapse attribution
\begin{equation}
\mathrm{SCPA}=1-\frac{\tau_{\mathrm{cond}}}{\tau_{\mathrm{auto}}}
=1-\frac{\theta_0}{\theta},
\end{equation}
which compares OU timescales before and after field conditioning. The second
is the model-free field-stripped persistence reduction, computed from the
e-folding lag or integrated ACF of $\psi_1$ versus that of
$\epsilon_{\mathrm{M2}}=\psi_1-\mu_{\mathrm{eff}}(v)$. These measures are not
expected to coincide exactly because they summarize different objects: one is
a rate parameter inside the fitted OU class, while the other is a residual
temporal statistic that remains sensitive to latent drivers, nonlinearity,
and rolling-window measurement effects \cite{White1982,Gardiner2009}. The
manuscript therefore interprets them jointly rather than forcing them into a
single percentage.

\section{Robustness Experiment Design and Execution}
\label{sec:app_robustness}

\subsection{Placebo-field protocol}
\label{sec:app_placebo_protocol}

The placebo experiment is executed as a fixed pipeline rather than as an
informal sensitivity check. First, we fit an AR($p$) model to the aligned
$\log(\mathrm{VIX})$ series, selecting $p$ by AIC over $p=1,\ldots,10$; this
chooses $p=9$ on the aligned sample. Second, we simulate 100 independent
surrogate paths of the same length as the empirical field, using the fitted
AR recursion and independent Gaussian innovations. Third, each path is
linearly rescaled to the empirical mean and variance of the real
$\log(\mathrm{VIX})$ series so that only the coupling information, not the
marginal scale, can differentiate the real and placebo fields. Fourth, each
surrogate replaces the real field in the exact M2 likelihood, and the
improvement over M0 is recorded as $\Delta$BIC$_{\mathrm{placebo}}$. Fifth,
the empirical $p$-value is computed as the fraction of placebo gains that meet
or exceed the real gain \cite{Theiler1992,White1982}. The same logic is reused
in the 2D daily placebo gate, where feedforward and full bidirectional models
are compared against the decoupled benchmark under persistence-matched placebo
drivers.

\subsection{Mechanical-proxy construction and standalone field tests}
\label{sec:app_mech_protocol}

The mechanical proxy is constructed on the same aligned daily sample used for
the main M2 fit. For each stock $i$, let
$s_{i,t}$ denote the 60-day rolling volatility and let
$\tilde{s}_i=\mathrm{median}_t\,s_{i,t}$ be its sample median over the aligned
period. With equal portfolio weights $w_i=1/N$, the mechanical portfolio
variance is
\begin{equation}
\widehat{V}^{\mathrm{mech}}_t
= \sum_{i=1}^{N}\sum_{j=1}^{N}
w_i w_j \tilde{s}_i \tilde{s}_j C_{ij}(t).
\end{equation}
The mechanical VIX proxy is then
\begin{equation}
\mathrm{VIX}_{\mathrm{mech},t}
= c\sqrt{\widehat{V}^{\mathrm{mech}}_t},
\end{equation}
where the scalar $c$ is chosen so that the proxy matches the sample mean of
the observed VIX. The informational component is defined by residualizing
$\log(\mathrm{VIX}_t)$ on $\log(\mathrm{VIX}_{\mathrm{mech},t})$. The design
then branches into two complementary evaluations. A static regression
decomposition yields the mechanical and informational fractions of shared
explanatory power. Separate M2 fits using the actual field, the mechanical
proxy, and the informational residual then ask which component carries
standalone model-selection power. The key inference is dynamic rather than
purely correlational: even if the mechanical proxy is highly correlated with
$\psi_1$, it does not count as explanatory unless it improves the stochastic
likelihood over M0.

To test recipe sensitivity, we repeat the construction under three
volatility-freezing statistics and three weighting schemes. The freeze
variants are the full-sample median, the full-sample mean, and the pre-2016
median of each stock's 60-day rolling volatility. The weighting variants are
equal weights, inverse-volatility weights, and volatility-share weights.
Every variant is pushed through the same two outputs: the static regression
decomposition and the standalone M2 fits with mechanical-only and
informational-only fields. This keeps the sensitivity analysis aligned to the
reviewer's actual concern, namely whether the qualitative model-selection
conclusion changes under reasonable alternative constructions.

\subsection{Quiet-regime pooling and field-stripped ACF evaluation}
\label{sec:app_acf_protocol}

The quiet-regime experiment is executed in two passes. The strict pass scans
the aligned daily sample for maximal contiguous segments in which the daily VIX
level remains inside $[15,18]$ for at least 120 consecutive trading days. When
that yields no valid segment, the fallback pass replaces the raw daily VIX by
a 20-day rolling median and scans three neighboring bands:
$[13,21]$, $[14,20]$, and $[15,19]$. For each band, all maximal contiguous
segments with length at least 120 trading days are retained. Within each
segment we compute the sample ACF of $\psi_1$, then pool the segment-specific
ACFs by available pair counts at each lag. The e-folding lag is reported as
the first lag at which the pooled ACF falls below $e^{-1}$. For the baseline
rolling-median band $[14,20]$, finite-sample uncertainty is quantified by
resampling the qualifying quiet episodes with replacement, which preserves
within-episode serial dependence, and recomputing the pooled e-folding lag
over 5000 bootstrap draws.

The field-stripped residual experiment uses the fitted M2 equilibrium path,
\begin{equation}
\epsilon_{\mathrm{M2}}(t)=\psi_1(t)-\mu_{\mathrm{eff}}(v_t),
\end{equation}
computed on the same aligned sample. We compare the e-folding lag and the
integrated ACF mass of raw $\psi_1$ and $\epsilon_{\mathrm{M2}}$ over fixed
horizons of 60 and 90 trading days. This pairing is intentional: the quiet
regime removes field variation by sample restriction, whereas the residual
experiment removes the fitted equilibrium path algebraically. Agreement
between the two strengthens the attribution claim because it shows persistence
collapse under two conceptually distinct manipulations.

\subsection{Window sweep, non-overlapping reconstructions, auxiliary fields, and out-of-sample sweeps}
\label{sec:app_window_protocol}

The window sweep is conducted by recomputing $\psi_1$ from raw daily returns at
five window lengths: $W=30,45,60,90,120$. No smoothing approximation is used;
each series is reconstructed from the returns and then aligned to the daily
VIX field over the corresponding date range. At each $W$, M0 and M2 are fit
from scratch and four outputs are retained:
$\tau_0$, $\tau_{\mathrm{cond}}$, $\beta/\theta$, and
$\mathrm{SCPA}=1-\tau_{\mathrm{cond}}/\tau_0$. This protocol cleanly
separates resolution dependence from code reuse, and it also provides an exact
check that the $W=60$ reconstruction reproduces the baseline order parameter.

To confront overlap directly, we also reconstruct disjoint weekly observables
from 5-trading-day correlation matrices sampled every 5 trading days. This is
done twice: once for a weekly $\psi_1$ proxy and once for weekly mean market
correlation, both aligned to weekly VIX closes
\cite{Wand2023Entropy,Wand2023Arxiv}. For a same-horizon boundary check, we
construct disjoint 60-trading-day blocks of $\psi_1$ and pair each block with
either its end-of-block VIX level or its within-block mean VIX. These
low-frequency series are intentionally retained even when underpowered,
because their role is to show how much of the core attribution survives when
measurement overlap is aggressively removed.

Auxiliary-field controls are run on the maximal same-sample intersections with
the respective fields. MOVE is tested by fitting an OU+MOVE model and then a
two-field OU+VIX+MOVE model on the intersection sample. TED is tested by
fitting an OU+TED model on its own maximal intersection sample. The same BIC
logic is used in every case. This design avoids misleading comparisons across
unequal date ranges while still showing whether non-VIX fields carry
independent dynamic information.

Out-of-sample evaluation is run at two levels. The baseline chronological
holdout uses a single split at 2016-01-01. Parameters are estimated on the
pre-2016 sample and scored on the post-2016 sample with the same exact
one-step Gaussian likelihood. We report the average log likelihood on train
and test and their ratio. Because the test set includes the COVID-19 period,
the holdout is meaningfully harder than a random split and therefore functions
as a basic stability stress test for the conditional field-proxy claim. As a sensitivity
check, we also exclude 2020-03-01 through 2020-09-30 from the test period and
score the pre- and post-COVID test segments separately, so that no one-step
increment is evaluated across the excluded gap.

To avoid hinging on one split, we then run an anchored sweep with cut dates at
2010-01-01, 2012-01-01, 2014-01-01, 2016-01-01, 2018-01-01, and 2020-01-01.
For each split, M0 and M2 are fit on the available prefix and scored on the
entire held-out suffix under the same exact one-step Gaussian likelihood. The
reported outputs are the M2 and M0 test log likelihoods per observation, the
M2$-$M0 test gap, and the M2 test/train ratio. This design does not convert
the problem into independent folds, but it does show whether the field-coupled
advantage survives repeated chronological re-estimation.

\subsection{Two-dimensional hidden-variable protocol and residual-state test}
\label{sec:app_twod_protocol}

The 2D hidden-variable protocol is run first on the daily aligned
$(\psi_1,\log\mathrm{VIX})$ series and then on two Wand-faithful weekly
reconstructions. In each dataset we fit three exact linear-Gaussian models:
decoupled, feedforward, and bidirectional. The comparison is based on exact
Gaussian likelihood and BIC. When the bidirectional model wins, we compute the
projected kernel amplitude and timescale implied by the fitted continuous-time
matrix. The stage gate is not whether any bidirectional model exists, but
whether the observed VIX yields both bidirectional preference and a projected
timescale compatible with the generalized-Langevin benchmark. That is why the
daily, weekly-thinning, and Wand-faithful reconstructions are all retained.

The residual-state test uses the orthogonalized level residual
$\epsilon_{\perp}$ rather than the mechanistic residual
$\epsilon_{\mathrm{M2}}$. After splitting the
$(\log\mathrm{VIX},\epsilon_{\perp})$ plane into four quadrants, we perform a
focused forecasting comparison between Q2 and Q3 because that contrast
captures the only genuinely operational claim proposed during manuscript
development: conditional on low VIX, does a positive orthogonal residual
predict a larger future increase in VIX than a negative one? We report mean
future changes, Mann--Whitney $p$-values, and rank-biserial effect sizes at
30, 60, and 90 days. The design is intentionally narrow so that the result is
either a clean positive or a clean negative rather than a diffuse narrative
about state-space geometry.

\section{Supplementary Quantitative Tables}
\label{sec:app_tables}

\begin{table}
\caption{\label{tab:window_robustness_appendix}Exact window-size robustness of
M0 and M2, recomputing $\psi_1$ directly from raw returns at each window.
SCPA denotes $1-\tau_{\mathrm{cond}}/\tau_{\mathrm{auto}}$.}
\begin{ruledtabular}
\begin{tabular}{ccccccccc}
$W$ & $\theta_0$ & $\tau_0$ & $\theta$ & $\tau_{\mathrm{cond}}$ & $\beta$ & $\beta/\theta$ & SCPA & $\Delta$BIC \\
\hline
30 & 0.00955 & 104.67 & 0.02927 & 34.17 & 0.00982 & 0.336 & 0.674 & 104.0 \\
45 & 0.00553 & 180.79 & 0.02093 & 47.78 & 0.00706 & 0.337 & 0.736 & 99.4 \\
60 & 0.00335 & 298.10 & 0.01640 & 60.99 & 0.00572 & 0.349 & 0.795 & 109.0 \\
90 & 0.00194 & 516.68 & 0.01317 & 75.93 & 0.00488 & 0.370 & 0.853 & 153.6 \\
120 & 0.00119 & 838.72 & 0.01036 & 96.55 & 0.00390 & 0.376 & 0.885 & 165.4 \\
\end{tabular}
\end{ruledtabular}
\end{table}

\begin{table}
\caption{\label{tab:field_components_appendix}M2 fits with actual,
mechanical-only, and informational-only fields on the decomposition sample.}
\begin{ruledtabular}
\begin{tabular}{lcccc}
Field & $\theta$ & $\tau_{\mathrm{cond}}$ & $\beta$ & $\Delta$BIC vs M0 (decomp.\ sample) \\
\hline
Actual $\log(\mathrm{VIX})$ & 0.01723 & 58.05 & 0.00623 & 108.6 \\
Mechanical-only proxy & 0.00393 & 254.18 & 0.00026 & -8.3 \\
Informational residual & 0.00850 & 117.67 & 0.00471 & 78.6 \\
\end{tabular}
\end{ruledtabular}
\end{table}

\begin{table}
\caption{\label{tab:decomp_sensitivity_appendix}Sensitivity of the
mechanical/informational decomposition to volatility-freezing and weighting
choices. The static split is recipe dependent, but the standalone dynamic
conclusion is invariant across all tested variants.}
\begin{ruledtabular}
\begin{tabular}{lcccc}
Recipe & Mechanical fraction & Informational fraction & $\Delta$BIC mech-only & $\Delta$BIC info-only \\
\hline
Median freeze, equal weights & 0.773 & 0.227 & -8.35 & 78.59 \\
Mean freeze, equal weights & 0.776 & 0.224 & -8.36 & 78.73 \\
Pre-2016 median freeze, equal weights & 0.779 & 0.221 & -8.34 & 78.62 \\
Median freeze, inverse-volatility weights & 0.815 & 0.185 & -8.27 & 77.71 \\
Median freeze, volatility-share weights & 0.721 & 0.279 & -8.38 & 79.57 \\
\end{tabular}
\end{ruledtabular}
\end{table}

\begin{table}
\caption{\label{tab:nonoverlap_appendix}One-dimensional non-overlapping
robustness checks. The disjoint weekly reconstructions preserve strong support
for M2, whereas same-horizon 60-day block reconstructions are retained only as
low-power boundary checks.}
\begin{ruledtabular}
\begin{tabular}{lccccc}
Construction & $n_{\mathrm{obs}}$ & $\Delta$BIC M2 vs M0 & $\tau_0$ (trading days) & $\tau_{\mathrm{cond}}$ (trading days) & SCPA \\
\hline
Daily overlapping baseline & 4973 & 109.0 & 298.10 & 60.99 & 0.795 \\
Disjoint weekly $\psi_1$ & 976 & 151.2 & 8.24 & 5.88 & 0.286 \\
Disjoint weekly mean correlation & 976 & 140.1 & 8.43 & 6.05 & 0.282 \\
Disjoint 60-day blocks, end-of-block VIX & 77 & 1.7 & 89.48 & 64.69 & 0.277 \\
Disjoint 60-day blocks, block-mean VIX & 77 & -3.4 & 89.48 & 75.05 & 0.161 \\
\end{tabular}
\end{ruledtabular}
\end{table}

\begin{table}
\caption{\label{tab:oos_splits_appendix}Anchored chronological out-of-sample
sweep. For each split date, M0 and M2 are fit on the pre-split prefix and
evaluated on the full held-out suffix.}
\begin{ruledtabular}
\begin{tabular}{lcccccc}
Split date & $n_{\mathrm{train}}$ & $n_{\mathrm{test}}$ & M0 test ll/obs & M2 test ll/obs & M2$-$M0 test gap & M2 test/train ratio \\
\hline
2010-01-01 & 1451 & 3522 & 3.1931 & 3.2000 & 0.0070 & 0.964 \\
2012-01-01 & 1955 & 3018 & 3.2227 & 3.2373 & 0.0146 & 0.994 \\
2014-01-01 & 2457 & 2516 & 3.2294 & 3.2437 & 0.0143 & 0.999 \\
2016-01-01 & 2961 & 2012 & 3.2219 & 3.2346 & 0.0128 & 0.995 \\
2018-01-01 & 3464 & 1509 & 3.1653 & 3.1805 & 0.0152 & 0.972 \\
2020-01-01 & 3967 & 1006 & 3.2301 & 3.2324 & 0.0023 & 0.995 \\
\end{tabular}
\end{ruledtabular}
\end{table}

\begin{table}
\caption{\label{tab:quiet_bands_appendix}Quiet-regime band sensitivity. The
strict daily band $[15,18]$ with a 120-trading-day minimum yields zero
qualifying segments over 2004--2023. For the baseline rolling-median band
$[14,20]$, the quiet e-folding estimate of 27 days carries an episode-bootstrap
$95\%$ confidence interval of $[18,33]$ days.}
\begin{ruledtabular}
\begin{tabular}{lccc}
Rolling-median VIX band & Segments $\ge 120$d & Quiet e-folding lag & Full-sample e-folding lag \\
\hline
$[13,21]$ & 5 & 27 & 69 \\
$[14,20]$ & 3 & 27 & 69 \\
$[15,19]$ & 0 & --- & 69 \\
\end{tabular}
\end{ruledtabular}
\end{table}

\begin{table}
\caption{\label{tab:rsvix_params_appendix}Parameters of the constrained
regime-switching plus VIX model M\_RS,c+VIX on the aligned daily sample.}
\begin{ruledtabular}
\begin{tabular}{lc}
Parameter or derived quantity & Estimate \\
\hline
$\theta$ & 0.02348 \\
$\mu_{\mathrm{calm}}$ & -0.6021 \\
$\mu_{\mathrm{stress}}$ & 1.0000 \\
$\beta$ & 0.00781 \\
$\sigma$ & 0.00818 \\
$p_{\mathrm{calm}\rightarrow\mathrm{stress}}$ & 0.01170 \\
$p_{\mathrm{stress}\rightarrow\mathrm{calm}}$ & 0.8940 \\
Expected calm duration (days) & 85.5 \\
Expected stress duration (days) & 1.12 \\
Stationary calm probability & 0.9871 \\
Stationary stress probability & 0.0129 \\
Calm-state relaxation time (days) & 42.6 \\
\end{tabular}
\end{ruledtabular}
\end{table}

\begin{table}
\caption{\label{tab:twod_appendix}Exact two-dimensional model comparisons and
projected-kernel timescales across the daily and Wand-faithful datasets. The
reported $\Delta$BIC values are relative to the winning model in each panel.}
\begin{ruledtabular}
\begin{tabular}{lcccc}
Dataset & Winning structure & $\Delta$BIC vs next-best & $\Delta$BIC vs decoupled & Kernel timescale (days) \\
\hline
Daily aligned $\psi_1$ & Feedforward & 2.44 & 125.26 & 36.46 (full model only) \\
Naive weekly thinning & Feedforward & 4.17 & 74.45 & 9.10 (full model only) \\
Wand-faithful weekly $\psi_1$ & Bidirectional & 0.78 & 154.10 & 33.47 \\
Wand-faithful weekly mean correlation & Bidirectional & 2.69 & 143.87 & 36.46 \\
\end{tabular}
\end{ruledtabular}
\end{table}

\bibliographystyle{apsrev4-2}
\bibliography{references}

\end{document}